\documentclass[pdftex,a4paper]{article}

\usepackage[a4paper]{geometry}

\usepackage[author-year]{amsrefs}
\usepackage{url}

\usepackage[utf8]{inputenc}        
\usepackage[english]{babel}

\usepackage{amsthm,amsmath,amsfonts,amssymb} 
\usepackage{amsthm}                
\usepackage{graphicx}
\usepackage{subfigure}               
                                    
\usepackage{listings}                
\usepackage{tikz}
\usepackage{algorithm2e}
\usepackage{mathtools}
\usepackage{enumerate}
\usepackage{authblk}
\usepackage{epstopdf}

\theoremstyle{plain}
\newtheorem{thm}{Theorem}[section]

\newtheorem{prop}[thm]{Proposition}

\theoremstyle{definition}

\newtheorem{exmp}{Example}[section]

\theoremstyle{remark}

\begin{document}
\title{Exploring a New Class of Non-stationary Spatial Gaussian Random Fields with Varying Local Anisotropy}
\author[1]{Geir-Arne Fuglstad\thanks{Corresponding author, fuglstad@math.ntnu.no}}
\author[2]{Finn Lindgren}
\author[1]{Daniel Simpson}
\author[1]{Håvard Rue}
\affil[1]{Department of Mathematical Sciences, NTNU, Norway}
\affil[2]{Department of Mathematical Sciences, University of Bath, UK}
\date{April 25, 2014}
\maketitle

\begin{abstract}
Gaussian random fields (GRFs) constitute an important part of spatial
modelling, but can be computationally infeasible for general
covariance structures. An efficient approach is to specify GRFs
via stochastic partial differential equations (SPDEs) and
derive Gaussian Markov random field (GMRF) approximations of the
solutions. We consider the construction of a class of non-stationary 
GRFs with varying local anisotropy, where the 
local anisotropy is introduced by allowing
the coefficients in the SPDE to vary with position. This is done by
using a form of diffusion equation driven by Gaussian white
noise with a spatially varying diffusion matrix. This allows
for the introduction of parameters that control the GRF by
parametrizing the diffusion matrix. These parameters and the GRF
may be considered to be part of a hierarchical model and the
parameters estimated in a Bayesian framework. The results show that the
use of an SPDE with non-constant coefficients is a promising way of
creating non-stationary spatial GMRFs that allow for physical
interpretability of the parameters, although there are several remaining 
challenges that would need to be solved before these models can be put 
to general practical use.

\end{abstract}

\section{Introduction}
Many spatial models for continuously indexed phenomena, such as temperature,
precipitation and air pollution, are based on Gaussian random fields (GRFs).
This is mainly due to the fact that their theoretical properties are
well understood and that their distributions can be fully
described by mean and covariance functions. In principle, it is enough to specify the mean at each location and the covariance between any two locations.
However, specifying covariance functions is hard and specifying covariance 
functions that can be controlled by parameters in useful ways is even harder. 
This is the reason why the covariance function usually is selected from a class of
known covariance functions such as the exponential covariance function, the 
Gaussian covariance function or the Mat\'{e}rn covariance function. 

But even when the covariance function is selected 
from one of these classes, the feasible
problem sizes are severely limited by a cubic increase in computation time
as a function of the number of observations and a quadratic
increase in computation time as a function of the number of prediction locations. 
This computational challenge is usually tackled
either by reducing the dimensionality of the 
problem~\cites{Cressie2008, Banerjee2008}, by introducing sparsity in the 
precision matrix~\cite{Rue2005} or the covariance matrix~\cite{Furrer2006},
or by using an approximate likelihood~\cites{Stein2004, Fuentes2007}. \ocite{Sun2012}
offers comparisons of the advantages and challenges associated with the usual
approaches to large spatial datasets.

The main goal of this paper is to explore a new class of
non-stationary GRFs that provide both an easy way to specify the parameters
and allows for fast computations. The main computational tool used is Gaussian
Markov random fields ({GMRFs})~\cite{Rue2005} with a spatial Markovian
structure where each position is conditionally dependent only on positions
close to itself. The strong connection between the Markovian structure
and the precision matrix results in sparse precision matrices that
can be exploited in computations.
The main problem associated with such an approach is that GMRFs must be 
constructed through conditional distributions, which presents a challenge
as it is generally not easy to determine whether a set of conditional 
distributions gives a valid joint distribution. Additionally, the 
conditional distributions have to be controlled by useful parameters in
such a way that not only the joint distribution is valid, but also such that
the effect of the 
parameters is understood. Lastly, it is desirable that the GMRF is 
a consistent approximation of a GRF
in the sense that when the distances between the positions decrease, the
GMRF ``approaches'' a continuous GRF. These issues are even more challenging
for non-stationary GMRFs. It is is extremely hard to specify the
non-stationarity directly through conditional distributions.

There is no generally accepted way to handle non-stationary GRFs, but many
approaches have been suggested. There is a large literature on methods based on
the deformation method of~\ocite{Sampson1992}, where a stationary process is
made non-stationary by deforming the space on which it is defined. Several
Bayesian extensions of the method have been 
proposed~\cites{Damian2001, Damian2003, Schmidt2003, Schmidt2011}, but all these methods require replicated realizations which might not be 
available. There has been some development
towards an approach for a single realization, but with a ``densely'' 
observed realization~\cite{Anderes2008}. Other approaches use
kernels which are convolved with Gaussian 
white noise~\cites{Higdon1998, Paciorek2006}, 
weighted sums of stationary processes~\cite{Fuentes2001} and expansions
into a basis such as a wavelet basis~\cite{Nychka2002}. 
Conceptually simpler methods have been made with
``stationary windows''~\cites{Haas1990a, Haas1990b} and with piecewise 
stationary Gaussian processes~\cite{Kim2005}. There has also been some progress
with methods based on the spectrum of the 
processes~\cites{Fuentes2001, Fuentes2002b, Fuentes2002a}. 
Recently, a new type of method based on a connection between stochastic
partial differential equations ({SPDEs}) and some classes
of GRFs was proposed by~\ocite{Lindgren2011}. They use an
SPDE to model the GRF and construct a GMRF approximation to the
GRF for computations. An application of a non-stationary model
of this type to ozone data can be found in~\ocite{Bolin2011} and 
an application to precipitation data can be found in~\ocite{Rikke2013}.

This paper extends on the work of~\ocite{Lindgren2011} and explores the possibility
of constructing a non-stationary GRF by varying the local anisotropy. 
The interest lies both in considering the different types of 
structures that can be achieved, and how to parametrize the GRF and 
estimate the parameters in a Bayesian setting. The construction of the GRF 
is based on an SPDE which describes the 
GRF as the result of a linear filter applied to Gaussian white noise. Basically,
the SPDE expresses how the smoothing of the Gaussian white noise varies at 
different locations. This construction bears some
resemblance to the deformation method of~\ocite{Sampson1992} in the sense that
parts of the spatial variation of the linear filter can be understood as
a local deformation of the space, only with an associated spatially varying
variance for the Gaussian white noise.
The main idea for  
computations is that since this filter works locally, it implies a Markovian
structure on the GRF. This Markovian structure can be transferred to a GMRF which
approximates the GRF, and in turn fast computations can be done with sparse matrices.

This paper presents a first look into a new type of model and the main goal is to 
explore what can be achieved in terms of models and inference with the model. 
Section~\ref{sec:Preliminaries} contains the motivation and introduction to the
class of non-stationary GRFs that is studied in the other sections. The form
of the SPDE that generates the class is given and it is related to more 
standard constructions of GMRFs. In Section~\ref{sec:Examples} illustrative
examples are given on both stationary and non-stationary constructions. This 
includes some discussion on how to control the non-stationarity of the GRF. Then 
Section~\ref{sec:Inference} explores parameter estimation for these types of models
through different examples with simulated data. The paper ends with discussion of 
extensions in Section~\ref{sec:Extensions} and general discussion
and concluding remarks in Section~\ref{sec:Discussion}.

\section{New class of non-stationary GRFs}
\label{sec:Preliminaries}
A GMRF \(\boldsymbol{u}\) is usually parametrized through a mean 
\(\boldsymbol{\mu}\) and a precision matrix \(\mathbf{Q}\) such that
\(\boldsymbol{u}\sim\mathcal{N}(\boldsymbol{\mu}, \mathbf{Q}^{-1})\). The main
advantage of this formulation compared to the usual parametrization of 
multivariate Gaussian distributions through
the covariance matrix is that the Markovian structure is represented in the non-zero
structure of the precision matrix \(\mathbf{Q}\)~\cite{Rue2005}. 
Off-diagonal entries are non-zero if and only if the corresponding elements of 
\(\boldsymbol{u}\) are conditionally independent. This can be seen
from the conditional properties of a GMRF,
\[
	\mathrm{E}(u_i | \boldsymbol{u}_{-i}) = \mu_i-\frac{1}{Q_{i,i}}\sum_{j\neq i} Q_{i,j}(u_j-\mu_j)
\]
and
\[
	\mathrm{Var}(u_i | \boldsymbol{u}_{-i}) = \frac{1}{Q_{i,i}},
\]
where \(\boldsymbol{u}_{-i}\) denotes the vector \(\boldsymbol{u}\) with
element \(i\) deleted. For a spatial GMRF the non-zeros of 
\(\mathbf{Q}\) can correspond to grid-cells that are close to 
each other in a grid, neighbouring regions in a Besag model and so on. 
However, even when this non-zero structure is determined it is not clear what 
values should be given to the non-zero elements of the precision matrix. 
This is the framework of the 
conditionally auto-regressive (CAR) models, whose conception predates
the advances in modern computational statistics~\cites{Whittle1954, Besag1974}.
In the multivariate Gaussian case it is clear that the requirement for
a valid joint distribution is that \(\mathbf{Q}\) is positive definite, which
is not an easy condition to check.

Specification of a GMRF through the conditional properties given above is
usually done in a somewhat ad-hoc manner. For regular grids, a process such as random
walk can be constructed and the only major issue is to get the conditional
variance correct as a function of step-length. For irregular grids the situation
is not as clear because each of the conditional means  and variances must 
depend on the varying step-lengths. In~\ocite{Lindgren2008} it is demonstrated that
some such constructions for second-order random walk can lead to
inconsistencies as new grid points are added, and they offer a surprisingly
simple construction for second-order random walk based on the SPDE
\[
	-\frac{\partial^2}{\partial x^2} u(x) = \sigma \mathcal{W}(x),
\]
where \(\sigma>0\) and \(\mathcal{W}\) is standard Gaussian white noise. 
If the precision matrix is chosen according to their scheme one does not
have to worry about scaling as the grid is refined, as it automatically
approaches the continuous second-order random walk. There is an
automatic procedure to select the form of the conditional means
and variances.

A one-dimensional second-order random walk is a relatively simple example 
of a process
with the same behaviour everywhere. To approximate a two-dimensional,
non-stationary GRF, a scheme would require (possibly) 
different anisotropy and correct
conditional variance at each location. To select the precision matrix in this
situation poses a large problem and there is abundant use of simple models
such as a spatial moving average
\[
	\mathrm{E}(u_{i,j}|\boldsymbol{u}_{-\{(i,j)\}}) = \frac{1}{4}(u_{i-1,j}+u_{i+1,j}+u_{i,j-1}+u_{i,j+1})
\]
with a constant conditional variance \(1/\alpha\). There are ad-hoc ways to extend
such a scheme to a situation with varying step-lengths in each direction, 
but little theory for more irregular choices of locations.

This is why the choice was made to start with the close connection between
SPDEs and some classes of GRFs that was presented in~\ocite{Lindgren2011}, 
which is not plagued by the issues above. 
From~\ocite{Whittle1954} it is known that the SPDE
\begin{equation}
	(\kappa^2-\Delta)u(\boldsymbol{s}) = \mathcal{W}(\boldsymbol{s}), \qquad \boldsymbol{s}\in\mathbb{R}^2,
	\label{eq:LindSPDE}
\end{equation}
where \(\kappa^2>0\) and \(\Delta = \frac{\partial^2}{\partial s_1^2}+\frac{\partial^2}{\partial s_2^2}\) is the Laplacian, gives rise to a GRF \(u\) with the
Mat\'{e}rn covariance function
\[
	r(\boldsymbol{s}) = \frac{1}{4\pi \kappa^2}(\kappa\lvert\lvert\boldsymbol{s}\rvert\rvert)K_1(\kappa\lvert\lvert\boldsymbol{s}\rvert\rvert),
\]
where \(K_1\) is the modified Bessel function of the second kind of order 1.
Equation~\eqref{eq:LindSPDE} can be extended to fractional operator orders 
in order to obtain other smoothness parameters in the Mat\'ern covariance
function.  However, for practical applications, the true smoothness of
the field is very hard to estimate from data, in particular when the
model is used in combination with an observation noise model.
Restricting the development to smoothness 1 in the Mat\'ern family is
therefore unlikely to be a major practical serious limitation.
However, for practical computations the model will be discretised
using methods similar to \ocite{Lindgren2011}, which does
permit other operator orders.  Integer orders are easiest, but for
stationary models, fractional orders are also achievable
\cite{Lindgren2011}*{Authors' discussion response}.  For
non-stationary models, techiques similar to \ocite{Bolin2013}*{Section 4.2} 
would be possible to use.  This means that even though we here
will restrict the model development to the special case in
Equation~\eqref{eq:LindSPDE}, other smoothnesses, e.g.\ exponential covariances, will
be reachable by combining the different approximation techniques.

The intriguing part, that~\ocite{Lindgren2011} expanded upon in 
Equation~\eqref{eq:LindSPDE}, is that
\((\kappa^2-\Delta)\) can be interpreted as a linear filter acting locally. This
means that if the continuously indexed process \(u\) were instead represented
by a GMRF \(\boldsymbol{u}\) on a grid or a triangulation, with appropriate
boundary conditions, one 
could replace this operator with
a matrix, say \(\mathbf{B}(\kappa^2)\), only involving neighbours of 
each location such that
Equation~\eqref{eq:LindSPDE} becomes approximately
\begin{equation}
	\mathbf{B}(\kappa^2)\boldsymbol{u} \sim \mathcal{N}(0,\mathbf{I}).
	\label{eq:MatrixEquation}
\end{equation}
The matrix \(\mathbf{B}(\kappa^2)\) depends on the chosen grid, but after the 
relationship is derived, the calculation of \(\mathbf{B}(\kappa^2)\) 
is straightforward for any \(\kappa^2\).
Since \(\mathbf{B}(\kappa^2)\) is sparse, the resulting precision matrix
\(\mathbf{Q}(\kappa^2) = \mathbf{B}(\kappa^2)^\mathrm{T}\mathbf{B}(\kappa^2)\) for 
\(\boldsymbol{u}\) 
is also sparse. This means that by 
correctly discretizing the operator (or linear filter), it
is possible to devise a GMRF with approximately the same distribution as the 
continuously indexed GRF. And because it comes from a continuous equation one
does not have to worry about changing behaviour as the grid is refined.

The class of models that are studied in this paper is the one that can be constructed
from Equation~\eqref{eq:LindSPDE}, but with anisotropy added to the \(\Delta\)
operator. A function \(\mathbf{H}\), that gives \(2\times 2\) symmetric
positive definite matrices at each position, is introduced and the operator
is changed to 
\begin{align*}
	\nabla\cdot\mathbf{H}(\boldsymbol{s})\nabla &=
	\frac{\partial}{\partial s_1}\left(h_{11}(\boldsymbol{s})\frac{\partial}{\partial s_1} \right)+
	\frac{\partial}{\partial s_1}\left(h_{12}(\boldsymbol{s})\frac{\partial}{\partial s_2} \right) \\ 
	&\phantom{=}+\frac{\partial}{\partial s_2}\left(h_{21}(\boldsymbol{s})\frac{\partial}{\partial s_1} \right)+
	\frac{\partial}{\partial s_2}\left(h_{22}(\boldsymbol{s})\frac{\partial}{\partial s_2} \right).
\end{align*}
This induces different strength of local dependence in different directions, 
which results in a range that varies with direction at all locations. Further, it is necessary for the discretization procedure to restrict the SPDE to a bounded domain. 
The chosen SPDE is
\begin{equation}
	(\kappa^2-\nabla\cdot\mathbf{H}(\boldsymbol{s})\nabla)u(\boldsymbol{s}) = \mathcal{W}(\boldsymbol{s}), \qquad \boldsymbol{s} \in \mathcal{D}=[A_1, B_1]\times[A_2, B_2]\subset\mathbb{R}^2,
	\label{eq:fullSPDE}
\end{equation}
where the rectangular domain makes it possible to use periodic
boundary conditions. Neither the rectangular shape of the domain nor
the periodic boundary conditions are essential restrictions for the
model, but are merely the practical restrictions we choose to work
with in this paper, in order to focus on the non-stationarity itself.

When using periodic boundary conditions when approximating the
likelihood of a stationary process on an unbounded domain, the
parameter estimates will be biased, e.g.\ when using the Whittle
likelihood in the two-dimensional case \cite{Dahlhaus1987edge}.
However, as \ocite{Lindgren2011}*{Appendix A.4} notes for the
case with Neumann boundary conditions (i.e.\ normal derivatives set to
zero), the effect of the boundary conditions is limited to a region in
the vicinity of the boundary.  At a distance greater than twice the
correlation range away from the boundary the bounded domain model is
nearly indistinguishable from the model on an unbounded domain.
Therefore, the bias due to boundary effects can be eliminated by
embedding the domain of interest into a larger region, in effect
moving the boundary away from where it would influence the likelihood
function.  For non-stationary models, defining appropriate boundary
conditions becomes part of the practical model formulation itself.
For simplicity we will therefore ignore this issue here, leaving
boundary specification for future development, but provide some
additional practical comments in Section~\ref{sec:Extensions}.

Both for interpretation and for the practical use of Equation~\eqref{eq:fullSPDE} 
it is useful to decompose \(\mathbf{H}\) into scalar functions. The anisotropy 
due to \(\mathbf{H}\) is decomposed as 
\[
	\mathbf{H}(\boldsymbol{s}) = \gamma\mathbf{I}_2 + \boldsymbol{v}(\boldsymbol{s})\boldsymbol{v}(\boldsymbol{s})^\mathrm{T},
\]
where \(\gamma\) specifies the isotropic, baseline effect and the vector field
\(\boldsymbol{v}(\boldsymbol{s}) = [v_x(\boldsymbol{s}), v_y(\boldsymbol{s})]^\mathrm{T}\)
specifies the direction and magnitude of the local, extra anisotropic effect
at each location. In this way, one can, loosely speaking, think of different Mat\'{e}rn
like fields locally each with its own anisotropy that are
combined into a full process. 
An example of an extreme case of a process with a strong local anisotropic effect
is shown in Example~\ref{exmp:nonStat}. The example
shows that there is a close connection between the vector field and the resulting
covariance structure of the GRF.

The main computational challenge is to determine the appropriate discretization 
of the SPDE in Equation~\eqref{eq:fullSPDE}, that is how to derive a
matrix \(\mathbf{B}\) such as in Equation~\eqref{eq:MatrixEquation}. 
The idea is to look to the field of numerics for discretization methods
for differential equations. Then combine these with properties of Gaussian
white noise. Namely, that for a Lebesgue measurable 
subset $A$ of $\mathbb{R}^n$, for some $n > 0$,
\[
        \int_A \! \mathcal{W}(\boldsymbol{s}) \, \mathrm{d}\boldsymbol{s} \sim \mathcal{N}(0,|A|),
\]
where $|A|$ is the Lebesgue measure of $A$, and that for two disjoint Lebesgue 
measurable subsets $A$ and $B$ of $\mathbb{R}^n$ the integral over $A$ and the 
integral over $B$ are independent~\cite{Adler2007}*{pp.~24--25}. 
A matrix equation such as Equation~\eqref{eq:MatrixEquation} 
was derived for the SPDE in Equation~\eqref{eq:fullSPDE} with a finite volume
method. The derivations are quite involved and technical and are in 
Appendix~\ref{app:DerivationPrecisionMatrix}. 
However, when the form of the discretized SPDE 
has been derived as an expression of the coefficients in the SPDE and the 
grid, the conversion from SPDE to GMRF is automatic for any choice of coefficients
and rectangular domain.

\section{Examples of models}
\label{sec:Examples}
The simplest case of Equation~\eqref{eq:fullSPDE} is with constant
coefficients. In this case one has an isotropic model (up to boundary
effects) if $\mathbf{H}$ is a constant times the identity matrix
or a stationary anisotropic model (up to boundary effects) if this
is not the case. In both cases it is possible to calculate
an exact expression for the covariance function and the marginal variance
for the corresponding SPDE solved over $\mathbb{R}^2$.

For this purpose write
\[
	\mathbf{H} = \begin{bmatrix} H_1 & H_2 \\ H_2 & H_3 \end{bmatrix},
\]
where $H_1$, $H_2$ and $H_3$ are constants. This gives the SPDE
\begin{equation}
\label{eq:genHomSPDE}
	\left[\kappa^2-H_1 \frac{\partial^2}{\partial x^2}-2H_2 \frac{\partial^2}{\partial x \partial y}-H_3 \frac{\partial^2}{\partial y^2}\right]u(\boldsymbol{s}) = \mathcal{W}(\boldsymbol{s}), \qquad \boldsymbol{s}\in\mathbb{R}^2.
\end{equation}
But if $\lambda_1$ and $\lambda_2$ are the eigenvalues of $\mathbf{H}$, then
the solution of the SPDE is actually only a rotated version of the solutions
of
\begin{equation}
\label{eq:rotHomSPDE}
	\left[\kappa^2-\lambda_1 \frac{\partial^2}{\partial \tilde{x}^2}-\lambda_2 \frac{\partial^2}{\partial \tilde{y}^2}\right]u(\boldsymbol{s}) = \mathcal{W}(\boldsymbol{s}), \qquad \boldsymbol{s}\in\mathbb{R}^2.
\end{equation}
Here the new $x$-axis is parallel to the eigenvector of $\mathbf{H}$ 
corresponding to $\lambda_1$ in the old coordinate system and
the new $y$-axis is parallel to the eigenvector of $\mathbf{H}$ 
corresponding to $\lambda_2$ in the old coordinate system.

From Proposition~\ref{prop:margVar} one can see that the marginal
variance of $u$ is
\[
	\sigma_m^2 = \frac{1}{4\pi\kappa^2\sqrt{\mathrm{det}(\mathbf{H})}} = \frac{1}{4\pi\kappa^2\sqrt{\lambda_1\lambda_2}}.
\]
One can think of the eigenvectors of $\mathbf{H}$ as the two principal
directions and $\lambda_1$ and $\lambda_2$ as a measure of the ``strength''
of the diffusion in these principal directions. Additionally, if 
$\lambda_1 = \lambda_2$, which is equivalent to $\mathbf{H}$ being
equal to a constant times the identity matrix, the SPDE is rotation
and translation invariant and the solution is isotropic. If 
$\lambda_1 \neq \lambda_2$, the SPDE is still translation invariant, but
not rotation invariant, and the solutions are stationary, but not
isotropic.

In our case the domain is not $\mathbb{R}^2$, but $[0,A]\times[0,B]$ with
periodic boundary conditions. This means that a boundary effect is introduced
and the above results are only approximately true.

\subsection{Stationary models}
For a constant \(\mathbf{H}\) the SPDE in Equation~\eqref{eq:fullSPDE} becomes
\[
	[\kappa^2-\nabla\cdot\mathbf{H}\nabla] u(\boldsymbol{s}) = \mathcal{W}(\boldsymbol{s}), \qquad \boldsymbol{s}\in[0,A]\times[0,B].
\]
This SPDE can be rewritten as
\begin{equation}
	[1-\nabla\cdot\hat{\mathbf{H}}\nabla] u(\boldsymbol{s}) = \sigma\mathcal{W}(\boldsymbol{s}), \qquad \boldsymbol{s}\in[0,A]\times[0,B],
	\label{eq:ReparSPDE}
\end{equation}
where \(\hat{\mathbf{H}} = \mathbf{H}/\kappa^2\) and \(\sigma = 1/\kappa^2\).
From this form it is clear that \(\sigma\) is only a scale parameter and that it is
enough to solve for \(\sigma=1\) and then multiply the solution with the desired
value of \(\sigma\). Therefore, it is the effect of \(\hat{\mathbf{H}}\) that
is most interesting to study.

It is useful to parametrize \(\hat{\mathbf{H}}\) as 
\[
	\hat{\mathbf{H}} = \gamma \mathbf{I}_2 + \beta \boldsymbol{v}(\theta)\boldsymbol{v}(\theta)^\mathrm{T},
\]
where $\boldsymbol{v}(\theta) = [\cos(\theta),\sin(\theta)]^\mathrm{T}$,
\(\gamma>0\) and \(\beta>0\). In this 
parametrization one can think of $\gamma$ as the coefficient of 
the second order derivative in the direction orthogonal to 
$\boldsymbol{v}(\theta)$ and $\gamma+\beta$ as the coefficient 
of the second order derivative in the direction $\boldsymbol{v}(\theta)$.
Ignoring boundary effects, $\gamma$ and $\gamma+\beta$ 
are the coefficients of the second order derivatives in 
Equation~\eqref{eq:rotHomSPDE} and $\theta$ is how much
the coordinate system has been rotated in positive direction.

\begin{exmp}[Stationary GMRF]
\label{exmp:statAni}
	The purpose of this example is to consider the effects of using a constant
	\(\hat{\mathbf{H}}\). Use the SPDE in Equation~\eqref{eq:ReparSPDE} with
	domain \([0,20]\times[0,20]\) and periodic boundary conditions, and
	discretize with a regular \(200\times 200\) grid. Two different values of
	\(\hat{\mathbf{H}}\) are used, an isotropic case with 
	\(\hat{\mathbf{H}} = \mathbf{I}_2\) and an anisotropic case with 
	\(\gamma = 1\), \(\beta=8\) and \(\theta = \pi/4\). The anisotropic case
	corresponds to a coefficient 9 in the \(x\)-direction and a coefficient
	1 in the \(y\)-direction, and then a rotation of \(\pi/4\) in the
	positive direction. The isotropic GMRF has marginal variances \(0.0802\)
	and the anisotropic GMRF has marginal variances \(0.0263\). For 
	comparison Proposition~\ref{prop:margVar} gives \(0.0796\) and
	\(0.0263\).

	Figure~\ref{fig:statAni-obs} shows one realization for each of the
	cases. Comparing Figure~\ref{fig:statAni-1-obs} and 
	Figure~\ref{fig:statAni-9-obs} it seems that the direction with the
	higher coefficient for the second-order derivative has longer
	range and more regular behaviour. Compared to the corresponding 
	partial differential equation (PDE) without the white noise, 
	this is what one would expect since 
	large values of the coefficient penalize large values of the second
	order derivatives. One should expect that the correlation range 
	increases when the coefficient is increased.

	\begin{figure}
		\centering
 		\subfigure{
 			\includegraphics[width=6cm]{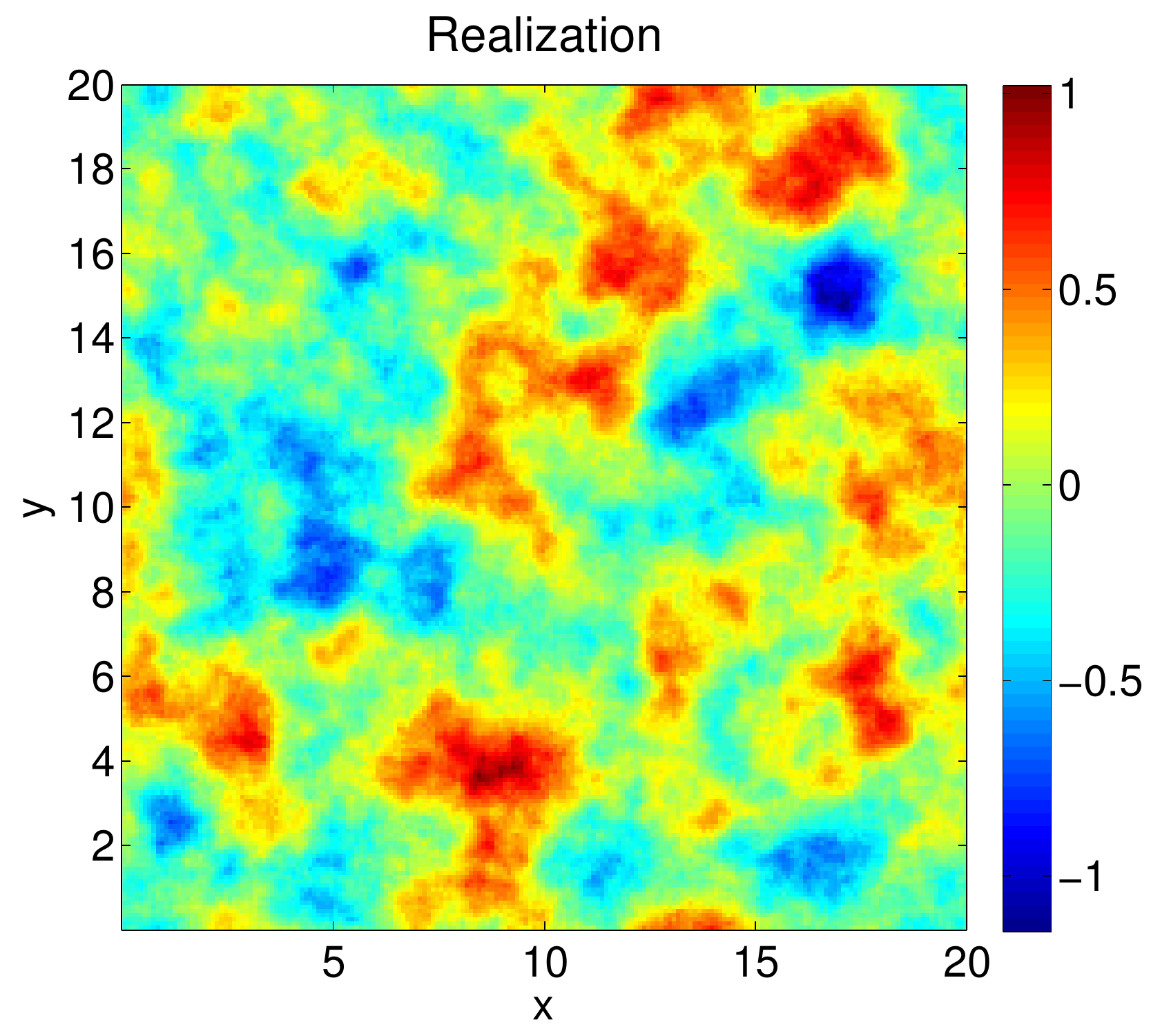}
 			\label{fig:statAni-1-obs}
 		}
 		\subfigure{
 			\includegraphics[width=6cm]{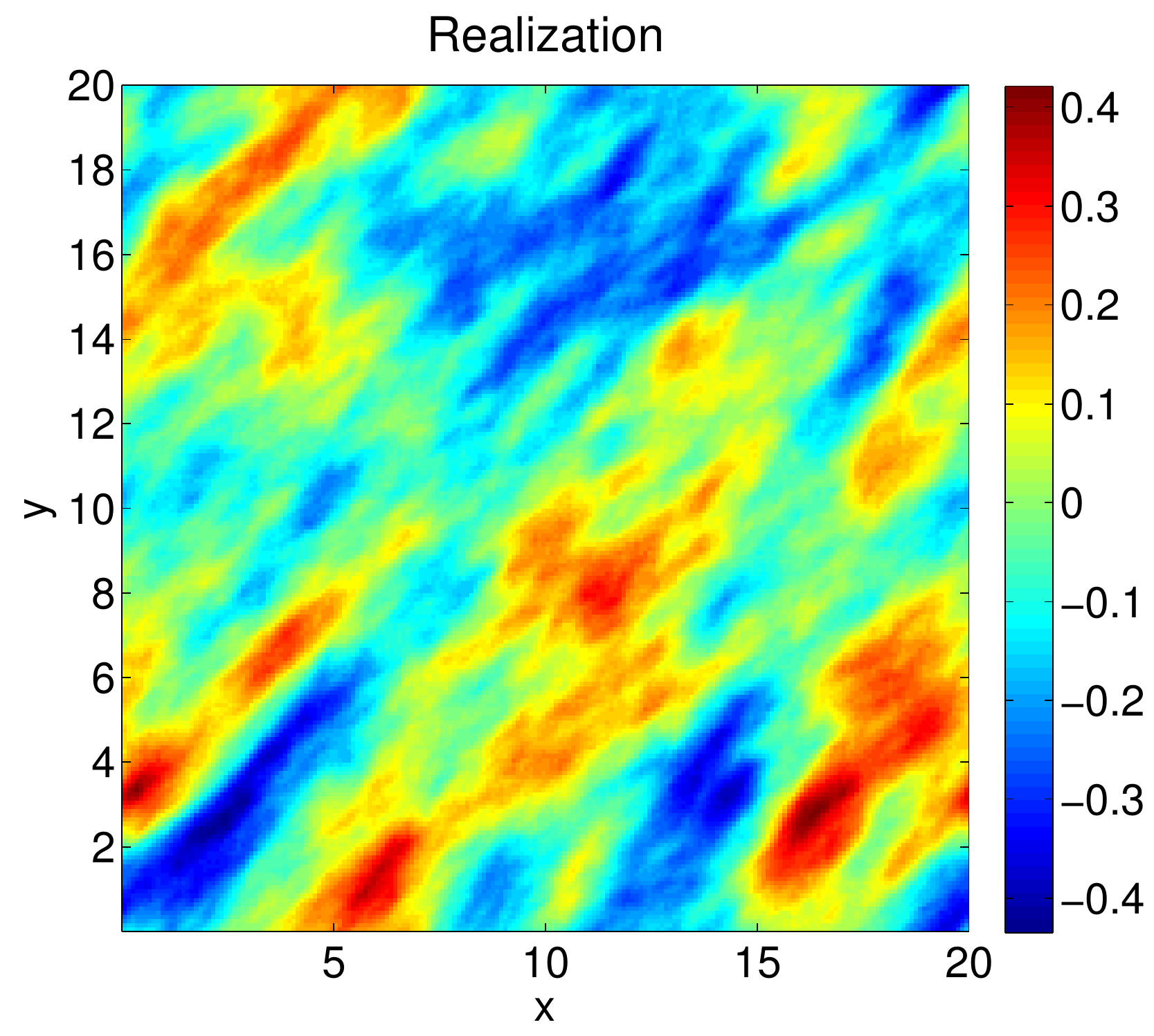}
 			\label{fig:statAni-9-obs}
 		}
 		\caption{\subref{fig:statAni-1-obs} 
 			Realization from the SPDE in Example~\ref{exmp:statAni}
 			on $[0,20]^2$ with a $200\times 200$ grid and periodic 
 			boundary conditions with $\gamma=1$, $\beta = 0$ and 
 			$\theta=0$. \subref{fig:statAni-9-obs} Realization 
 			from the SPDE in Example~\ref{exmp:statAni} on 
 			$[0,20]^2$ with a $200 \times 200$ grid and periodic 
 			boundary conditions with $\gamma = 1$, $\beta = 8$ and
 			$\theta = \pi/4$.}
 		\label{fig:statAni-obs}
 	\end{figure}

 	This is in fact what happens. Figure~\ref{fig:statAni-cov} shows the
 	correlation of the variable at \((9.95, 9.95)\) with every other
 	point in the grid for the isotropic and the anisotropic case. This is
 	sufficient to describe all the correlations since the solutions are
 	stationary. One can immediately note that the iso-correlation curves
 	are close to ellipses with semi-axes along \(\boldsymbol{v}(\theta)\)
 	and the direction orthogonal to \(\boldsymbol{v}(\theta)\). One can see that the
 	correlation decreases most slowly and most quickly in the directions
 	used to specify \(\hat{\mathbf{H}}\), with slowest decrease along 
 	\(\boldsymbol{v}(\theta)\). It is interesting to see
 	that both the isotropic case and the non-isotropic case has approximately
 	the same length for the minor semi-axis of the iso-correlation curves, and
 	that the major semi-axis is longer for the anisotropic case. This is
 	due to the fact that the lengths of the semi-axes are connected with
 	\(\sqrt{\gamma}\) and \(\sqrt{\gamma+\beta}\).

 	\begin{figure}
 		\centering
 		\subfigure{
 			\includegraphics[width=6cm]{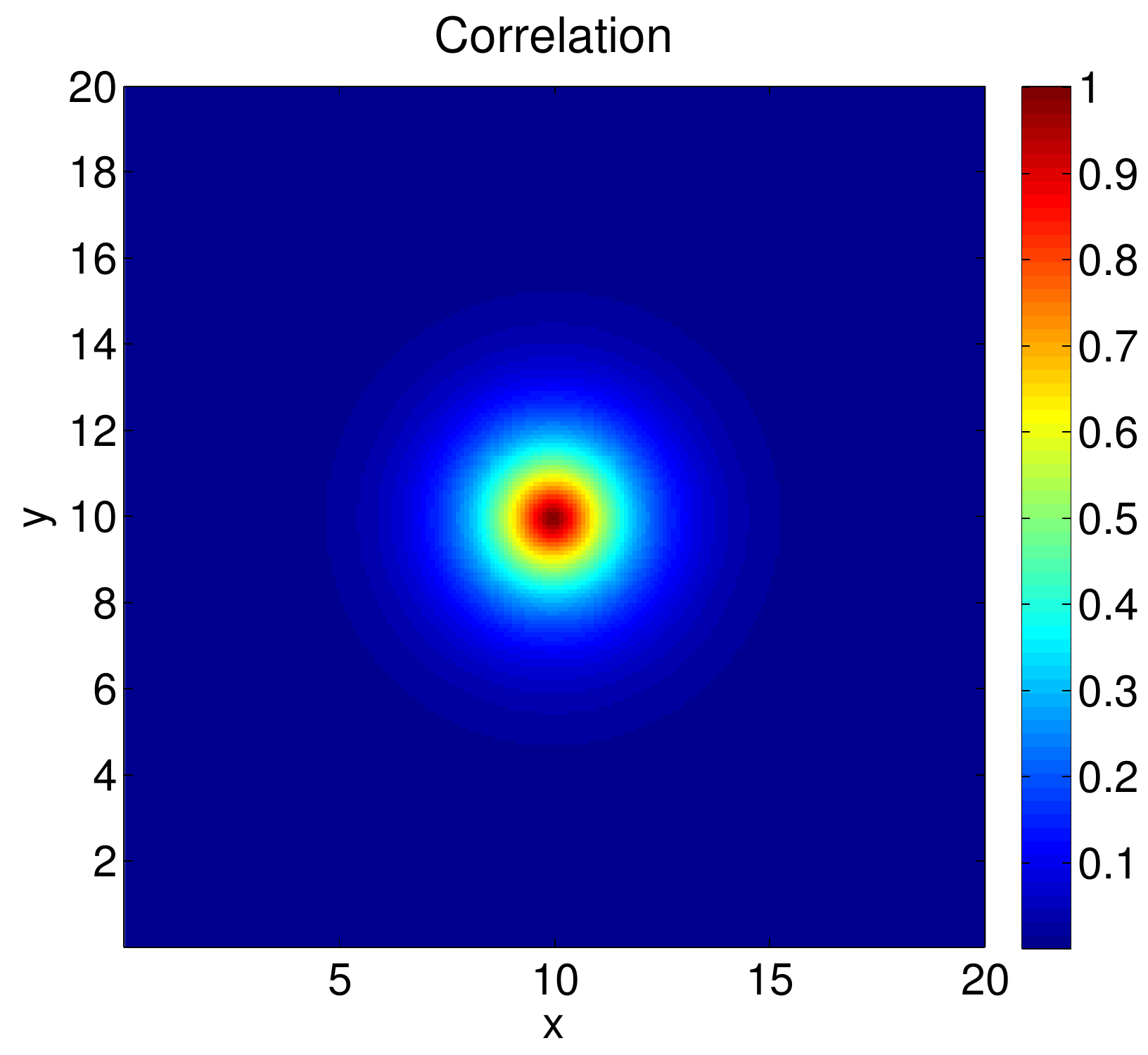}
 			\label{fig:statAni-1-cov}
 		}
 		\subfigure{
 			\includegraphics[width=6cm]{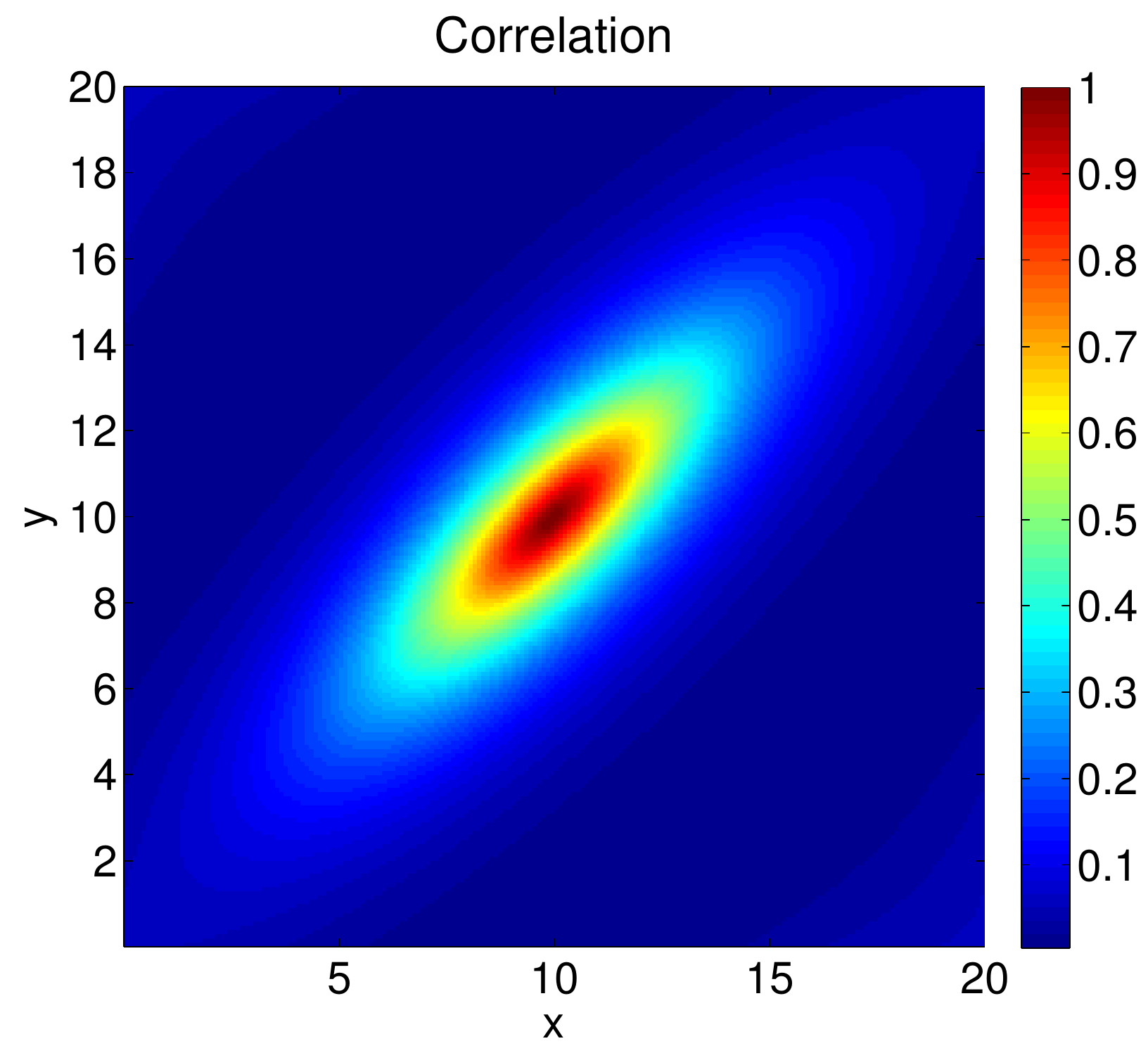}
 			\label{fig:statAni-9-cov}
 		}
 		\caption{\subref{fig:statAni-1-cov} 
 			Correlation of the centre with all other points for 
 			the solution of the SPDE in Example~\ref{exmp:statAni}
 			on $[0,20]^2$ with a $200\times 200$ grid and periodic 
 			boundary conditions with $\gamma=1$, $\beta = 0$ and 
 			$\theta = 0$. \subref{fig:statAni-9-cov}
 			Correlation of the centre with all other points for 
 			the SPDE in Example~\ref{exmp:statAni} on $[0,20]^2$ 
 			with a $200 \times 200$ grid and periodic boundary 
 			conditions with $\gamma = 1$, $\beta = 8$, 
 			$\theta=\pi/4$.}
 		\label{fig:statAni-cov}
 	\end{figure}

\end{exmp}

From the example above one can see that the use of 3 parameters allow for
the creation of GMRFs which are more regular in one direction than the other.
One can use the parameters \(\gamma\), \(\beta\) and \(\theta\) to control
the form of the correlation function and \(\sigma\) to get
the desired marginal variance.

\subsection{Non-stationary models}
To make the solution of the SPDE in Equation~\eqref{eq:fullSPDE} 
non-stationary, either $\kappa^2$ or $\mathbf{H}$ has to be a non-constant 
function. One way to achieve non-stationarity is by choosing 
\[
	\mathbf{H}(\boldsymbol{s}) = \gamma \mathbf{I}_2 + \beta \boldsymbol{v}(\boldsymbol{s})\boldsymbol{v}(\boldsymbol{s})^\mathrm{T},
\]
where $\boldsymbol{v}$ is a non-constant vector field on $[0,A]\times[0,B]$ 
which satisfy the periodic boundary conditions and $\gamma > 0$ and 
$\beta > 0$ are constants.

\begin{exmp}[Non-stationary GMRF]
\label{exmp:nonStat}
	Use the domain $[0,20]^2$ with a $200\times 200$ grid and
	periodic boundary conditions for the SPDE in 
	Equation~\eqref{eq:fullSPDE}. Let $\kappa^2$ be equal to 
	$1$ and let ${\sf \bf H}$ be given as
	\[
		\mathbf{H}(\boldsymbol{s}) = \gamma \mathbf{I}_2 + \beta \boldsymbol{v}(\boldsymbol{s})\boldsymbol{v}(\boldsymbol{s})^\mathrm{T},
	\]
	where $\boldsymbol{v}$ is a $2$-dimensional vector field on $[0,20]^2$ 
	which satisfies the periodic boundary conditions and $\gamma > 0$ and 
	$\beta > 0$ are constants. 

	To create an interesting vector field, start with the function 
	$f:[0,20]^2\rightarrow\mathbb{R}$ defined by
	\[
		f(x,y) = \left(\frac{10}{\pi}\right) \left(\frac{3}{4}\sin(2\pi x/20)+\frac{1}{4}\sin(2\pi y/20)\right).
	\]
	Then calculate the gradient $\nabla f$ and let 
	$\boldsymbol{v}:[0,20]^2\rightarrow\mathbb{R}^2$ be the gradient 
	rotated $90^{\circ}$ counter-clockwise at each point. 
	Figure~\ref{fig:nonStat-fvalues} shows the values of the function $f$ 
	and Figure~\ref{fig:nonStat-ffield} shows the resulting vector field 
	$\boldsymbol{v}$. The vector field is calculated on a $400 \times 400$ 
	regular grid, because the values between neighbouring cells in the 
	discretization are needed.

	\begin{figure}
		\centering
		\subfigure[The function used to create the vector field.]{
			\includegraphics[width=6cm]{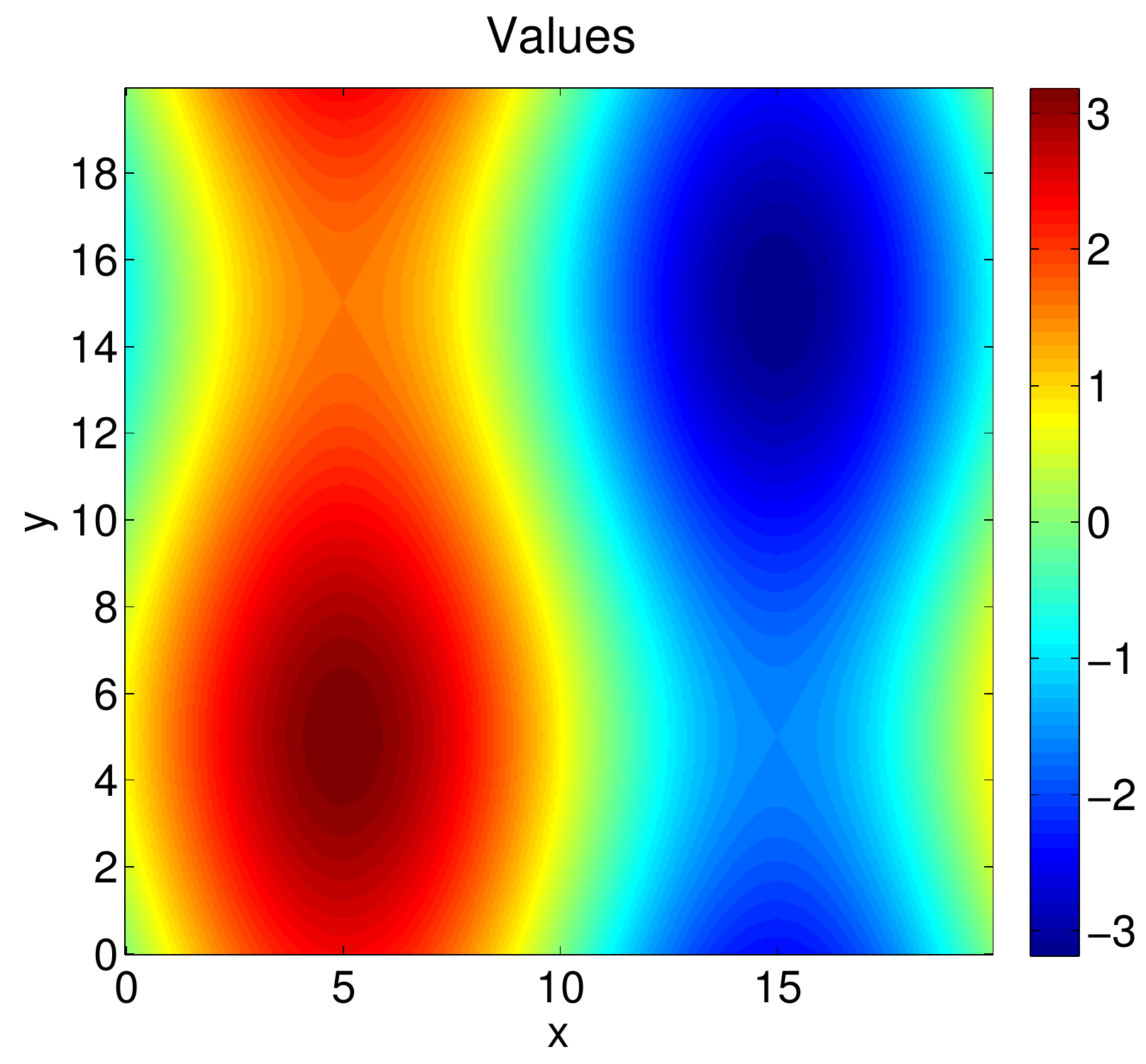}
			\label{fig:nonStat-fvalues}
		}
		\subfigure[The resulting vector field.]{
			\includegraphics[width=5.2cm]{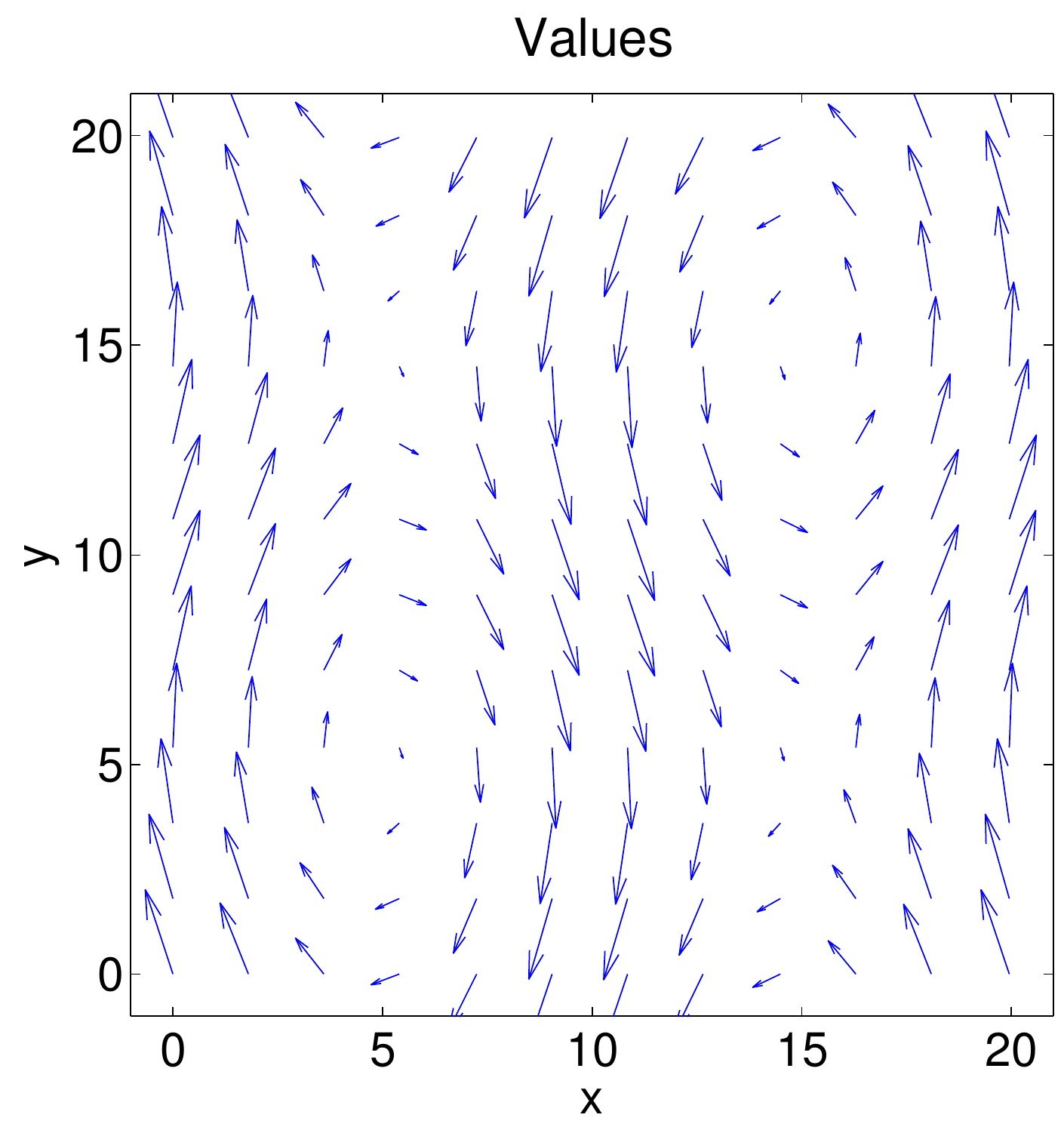}
			\label{fig:nonStat-ffield}
		}
		\caption{The gradient of the function illustrated 
			 in~\subref{fig:nonStat-fvalues} is calculated and 
			 rotated $90^{\circ}$ counter-clockwise at each point 
			 to give the vector field illustrated 
			 in~\subref{fig:nonStat-ffield}.}
		\label{fig:nonStat-fill}
	\end{figure}

	Figure~\ref{fig:nonStat-obs} shows one realization from the resulting 
	GMRF with $\gamma = 0.1$ and $\beta =25$. A much higher value for 
	$\beta$ than $\gamma$ is chosen to illustrate the connection between 
	the vector field and the resulting covariance structure. From the 
	realization it is clear that there is stronger dependence along the 
	directions of the vector field shown in 
	Figure~\ref{fig:nonStat-ffield} at each point than in the other 
	directions. In addition, from Figure~\ref{fig:nonStat-mvar} it seems 
	that positions with large values for the norm of the vector field has 
	smaller marginal variance than positions with small values and vice 
	versa. This feature introduces an undesired connection between
	anisotropy and marginal variances. It is possible to reduce this interaction
	between the vector field and the marginal variances by 
	reformulating the controlling SPDE as discussed briefly 
	in Section~\ref{sec:Extensions}.

	\begin{figure}
		\centering
		\subfigure[One realization.]{
			\includegraphics[width=6cm]{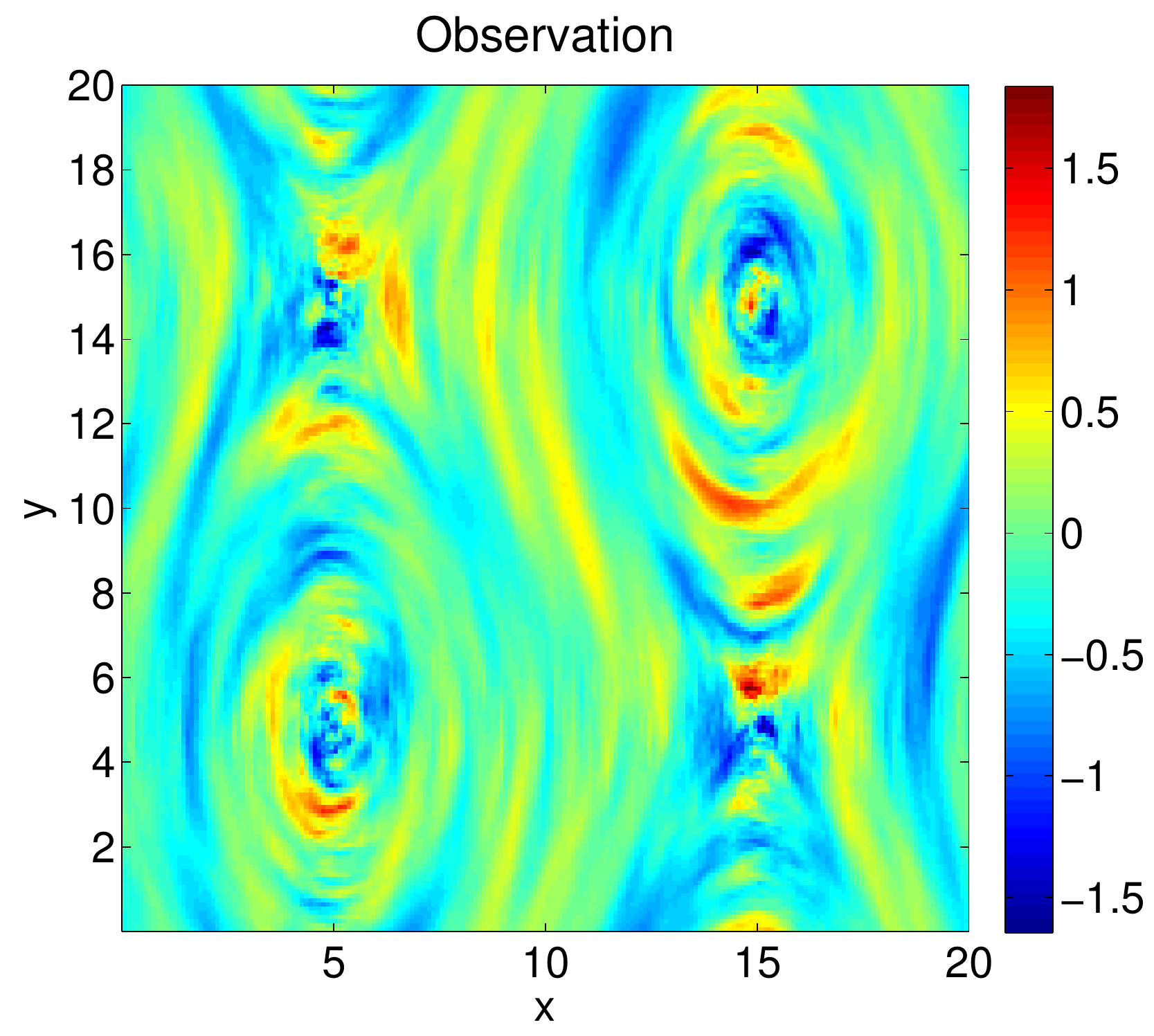}
			\label{fig:nonStat-obs}
		}
		\subfigure[Marginal variances.]{
			\includegraphics[width=6cm]{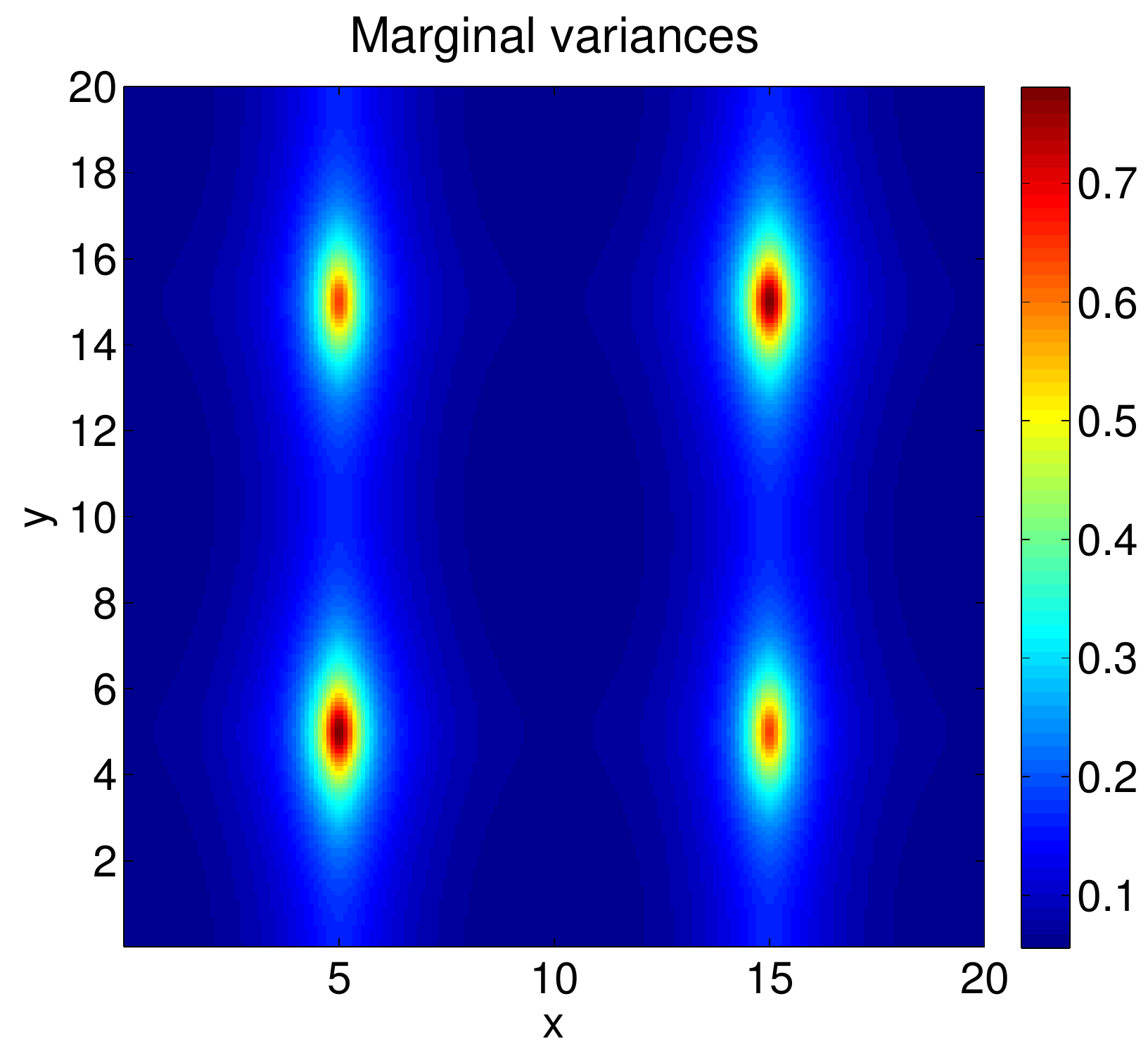}
			\label{fig:nonStat-mvar}
		}
		\caption{One observation and the marginal variances of the 
			 solution of the SPDE in Equation~\eqref{eq:fullSPDE} 
			 on a $200 \times 200$ 	regular grid of $[0,20]^2$ 
			 with periodic boundary conditions, 
			 $\kappa^2 \equiv 1$ and $\mathbf{H} = 0.1\mathbf{I}_2+25 \boldsymbol{v}\boldsymbol{v}^\mathrm{T}$, 
			 where $\boldsymbol{v}$ is the vector field described 
			 in Example~\ref{exmp:nonStat}.}
		\label{fig:nonStat-fobs}
	\end{figure}

	From Figure~\ref{fig:nonStat-covmm} and Figure~\ref{fig:nonStat-covs} 
	one can see that the correlations depend on the direction and norm 
	of the vector field, and that there is clearly non-stationarity. 
	Figure~\ref{fig:nonStat-covld} and Figure~\ref{fig:nonStat-covlu}
	show that the correlations with the positions $(4.95,1.95)$ 
	$(4.95,7.95)$ tend to follow the vector field around the point
	$(5,5)$, whereas Figure~\ref{fig:nonStat-covrd} and 
	Figure~\ref{fig:nonStat-covru} show that the correlations with
	the positions $(14.95,1.95)$ and $(14.95,7.95)$ tend to
	follow the vector field away from the point $(15,5)$.
	Figure~\ref{fig:nonStat-covlm} shows that the correlations with
	position $(4.95, 4.95)$ and every other point is not isotropic, but
	concentrated close to the point itself, and Figure~\ref{fig:nonStat-covrm}
	shows that the correlations with position $(14.95,4.95)$ have
	high correlation along four directions which extends out from the
	point. Figure~\ref{fig:nonStat-covmm} shows that the
	correlations with position \((9.95, 9.95)\) ``follow'' the vector field 
	with high correlations in the vertical direction.

	\begin{figure}
		\centering
		\includegraphics[width=6cm]{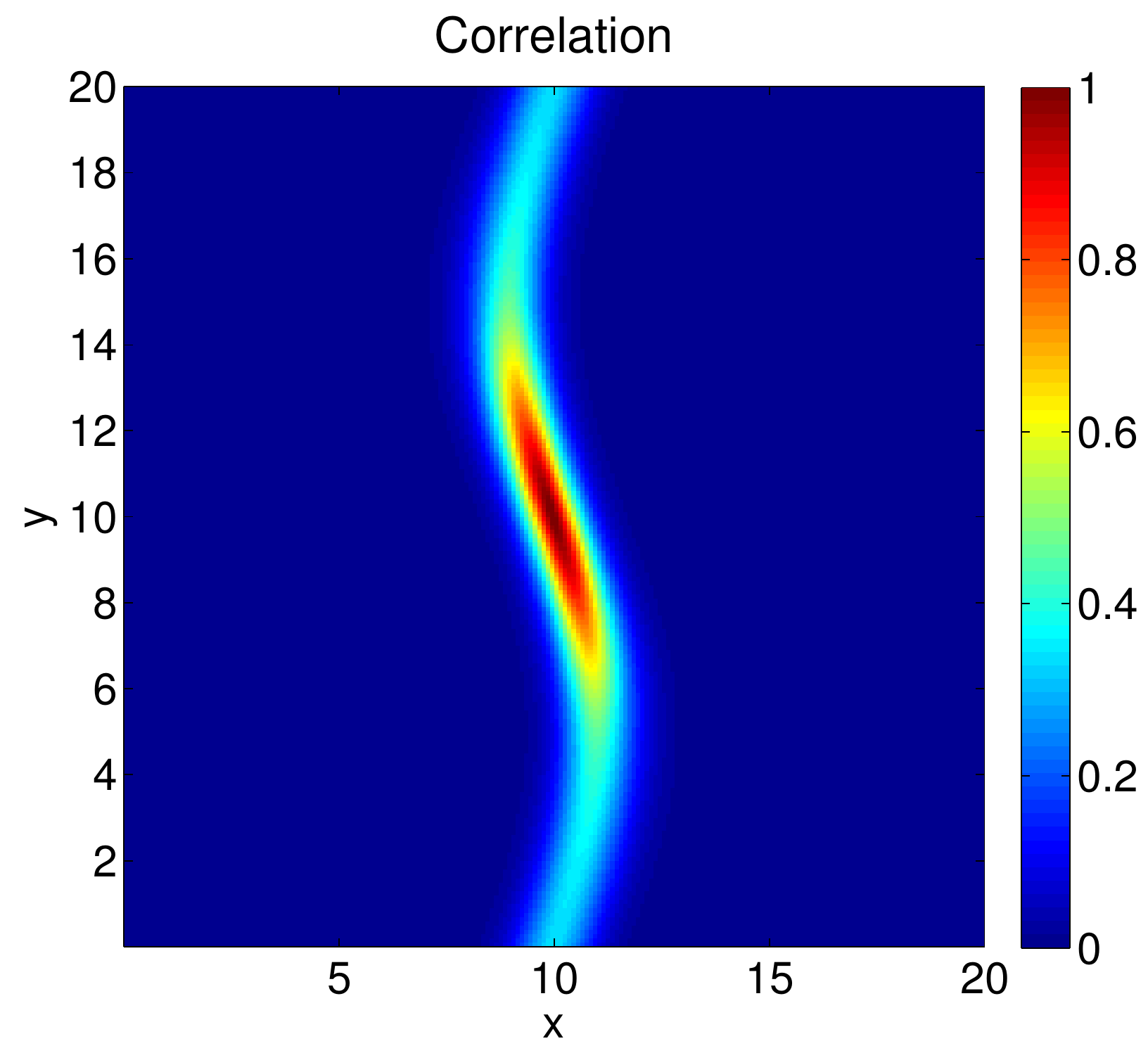}
		\caption{Correlations with position $(9.95,9.95)$ and all 
			 other points for the solution of the SPDE in
			 Example~\ref{exmp:nonStat}.}
		\label{fig:nonStat-covmm}
	\end{figure}
	
	\begin{figure}
		\centering
		\subfigure[Correlations with position $(4.95,1.95)$.]{
			\includegraphics[width=6cm]{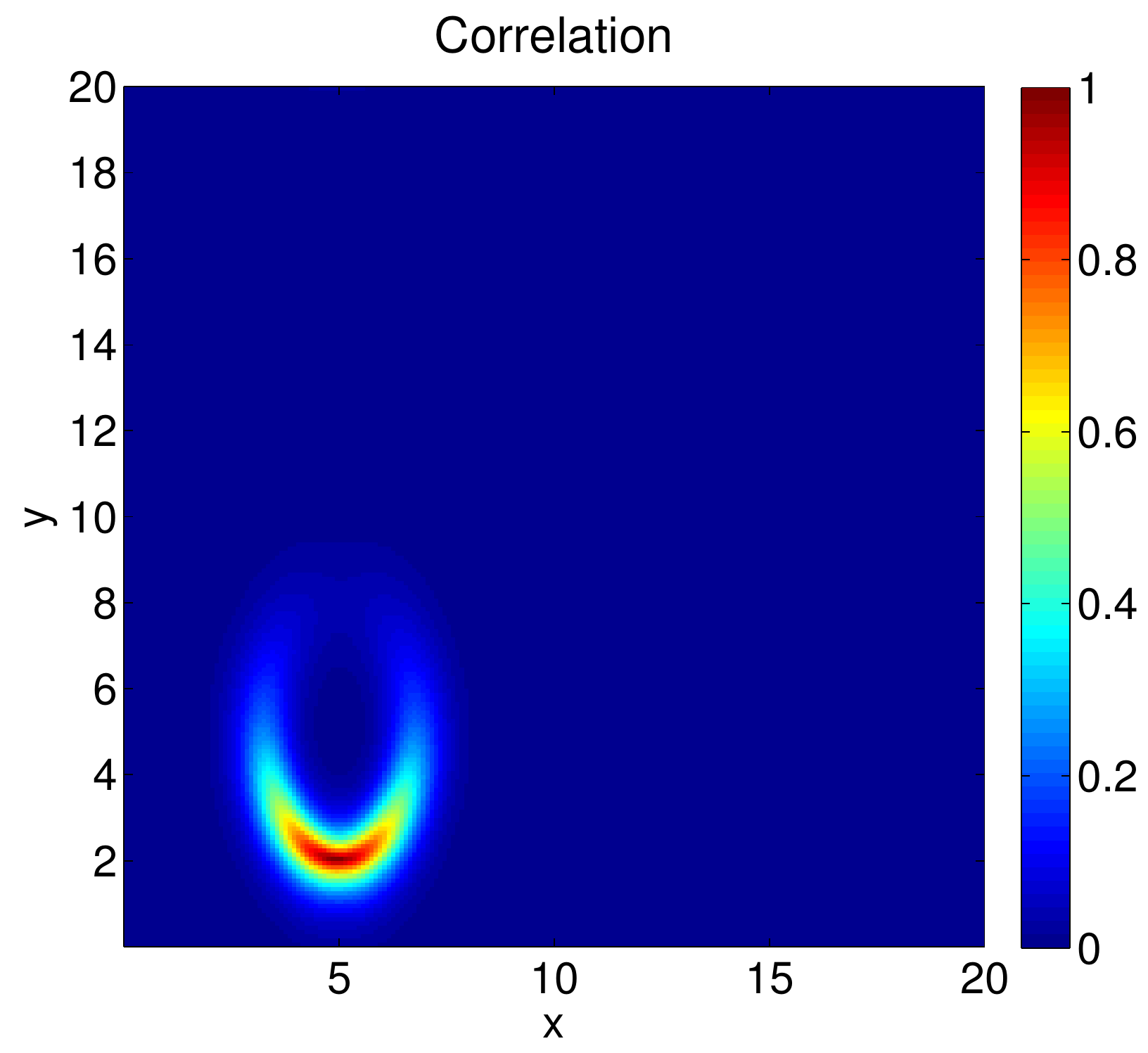}
			\label{fig:nonStat-covld}
		}
		\subfigure[Correlations with position $(14.95,2.05)$.]{
			\includegraphics[width=6cm]{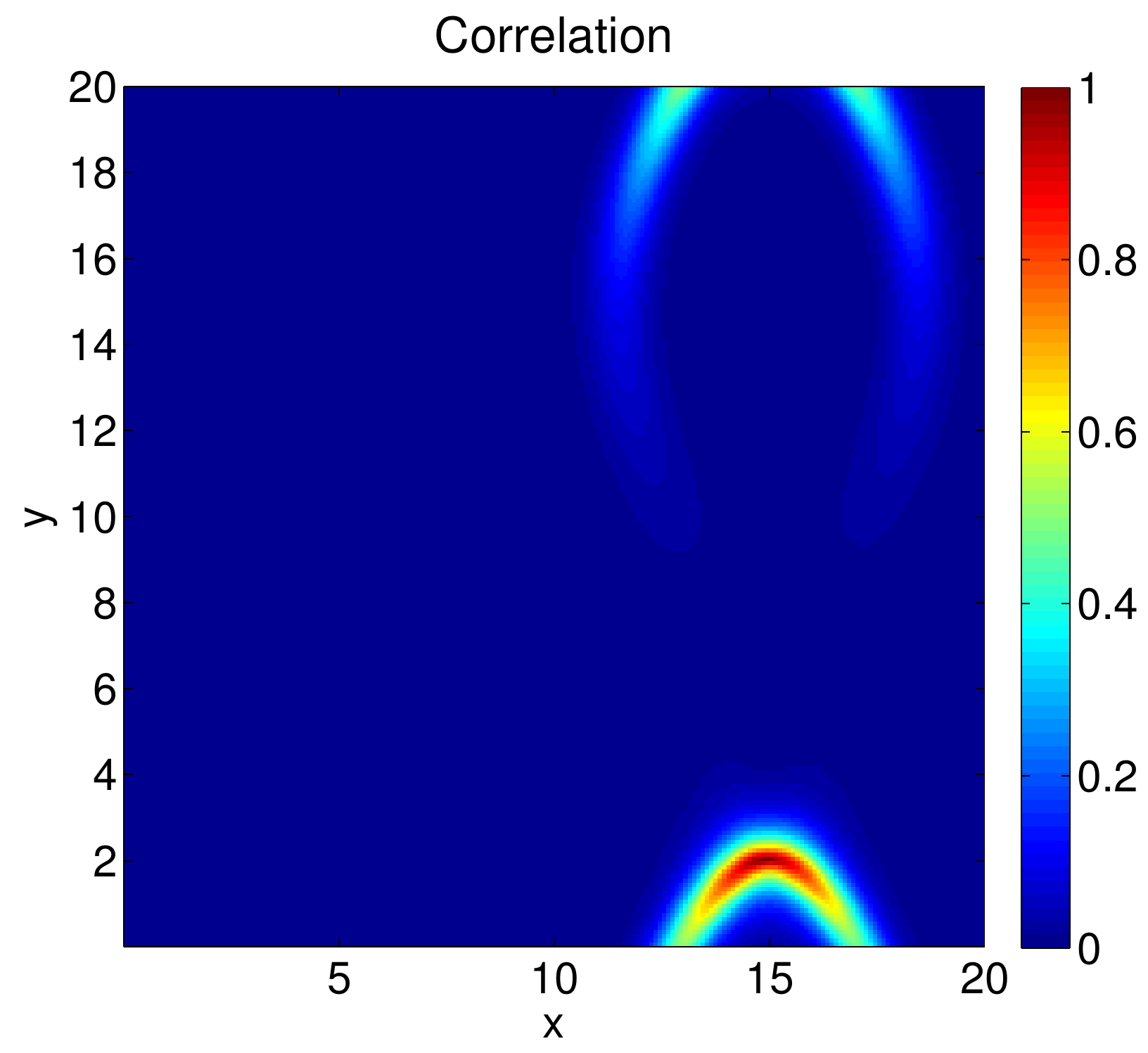}
			\label{fig:nonStat-covrd}
		}\\
		\subfigure[Correlations with position $(4.95, 7.95)$.]{
			\includegraphics[width=6cm]{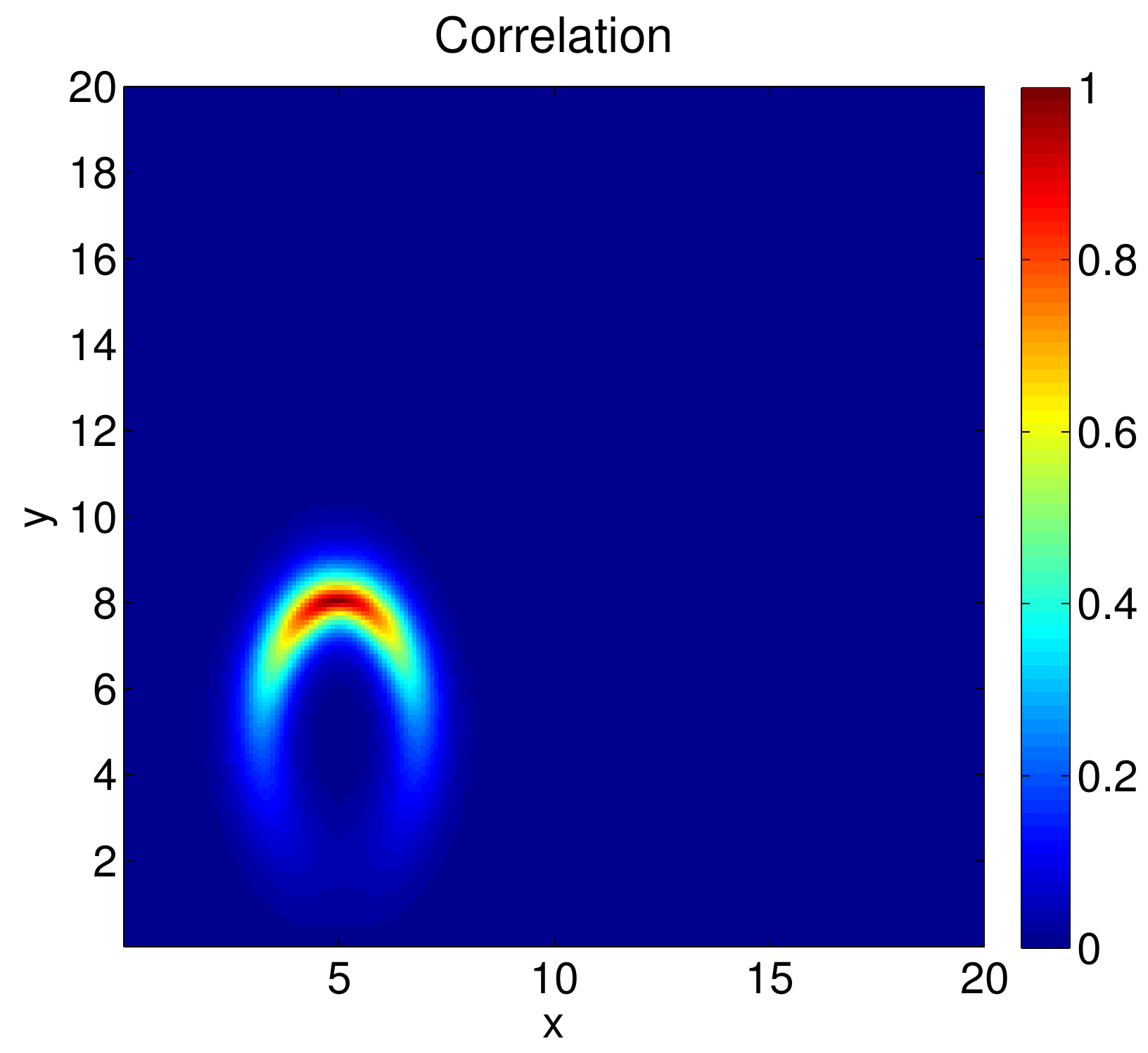}
			\label{fig:nonStat-covlu}
		}
		\subfigure[Correlations with position $(14.95, 7.95)$.]{
			\includegraphics[width=6cm]{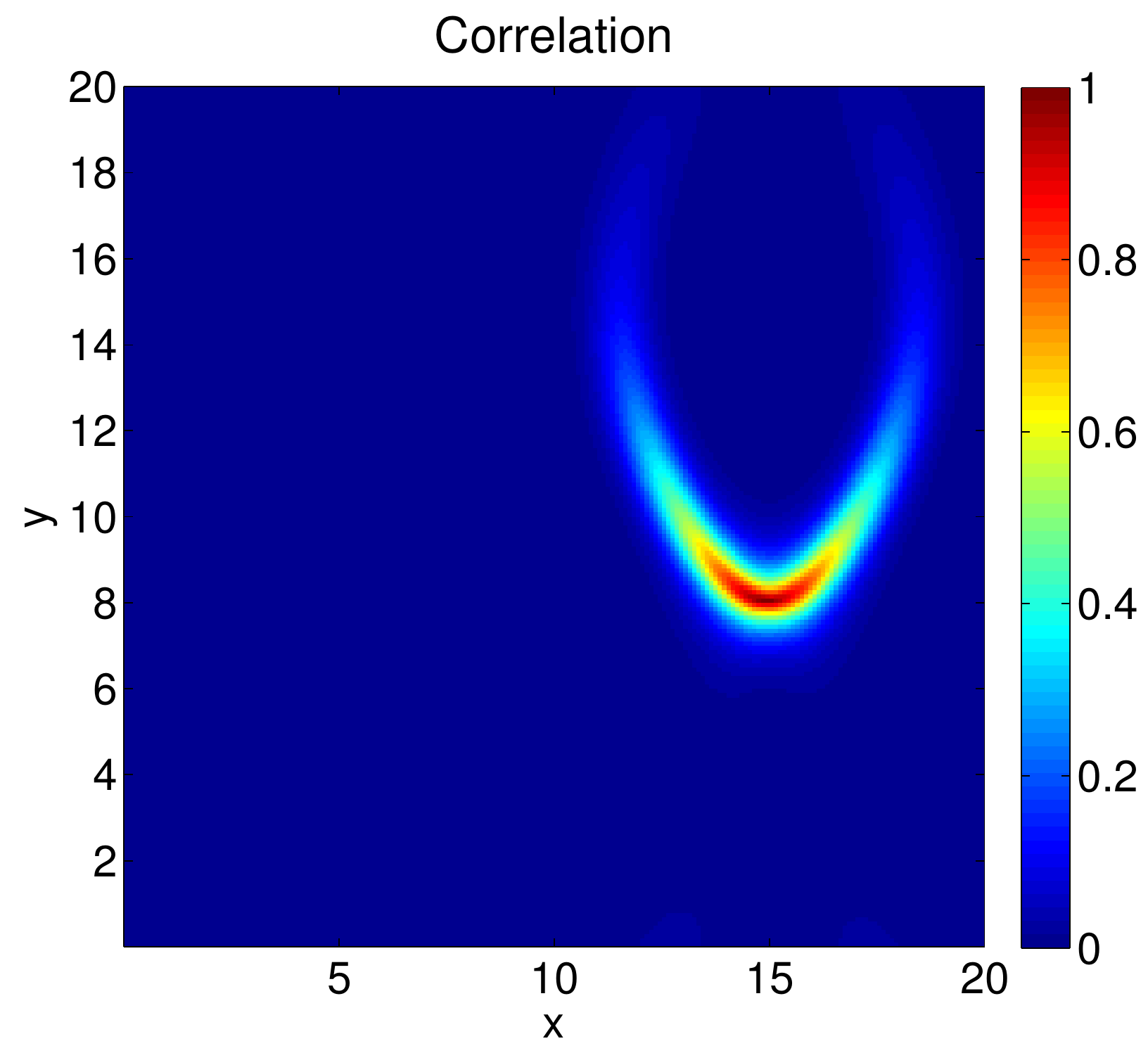}
			\label{fig:nonStat-covru}
		}\\
		\subfigure[Correlations with position $(4.95, 4.95)$.]{
			\includegraphics[width=6cm]{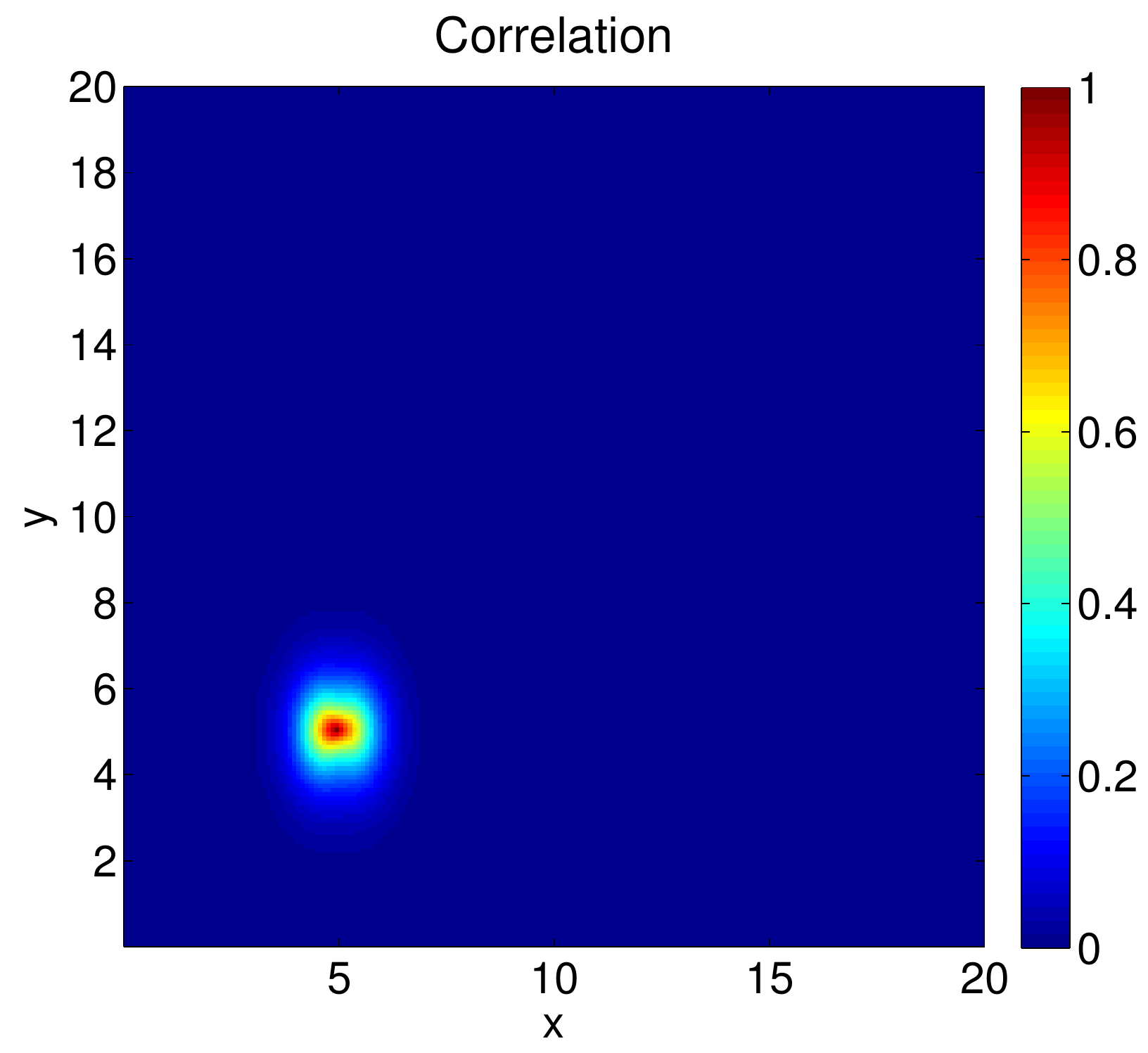}
			\label{fig:nonStat-covlm}	
		}
		\subfigure[Correlations with position $(14.95, 4.95)$.]{
			\includegraphics[width=6cm]{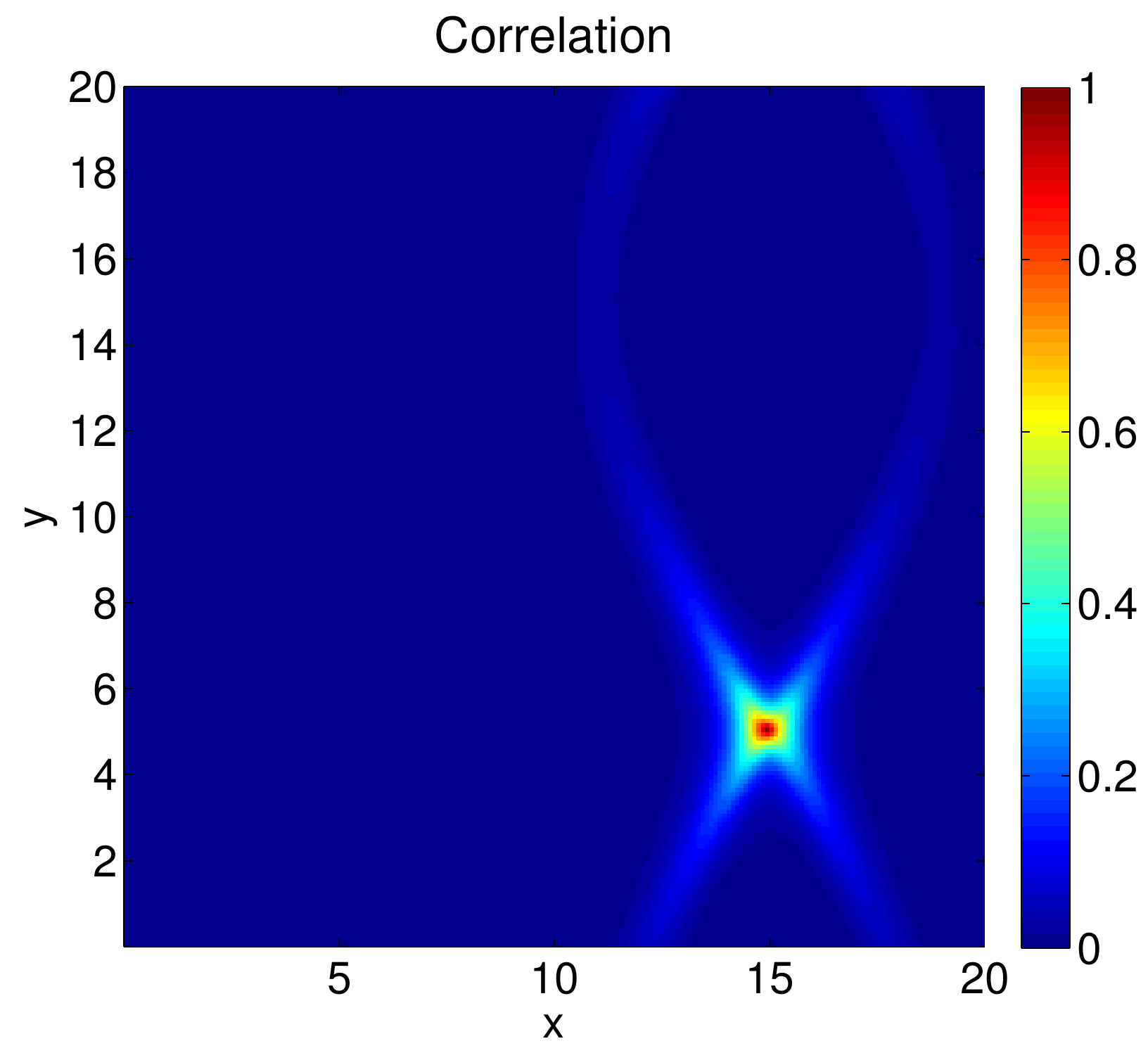}
			\label{fig:nonStat-covrm}
		}
		\caption{Correlations for different points with all other points for
			 the solution of the SPDE in Example~\ref{exmp:nonStat}.}
		\label{fig:nonStat-covs}
	\end{figure}
\end{exmp}

From this example one can see that allowing  $\mathbf{H}$ to be non-constant 
means that one can vary the dependence structure in more interesting ways than 
the stationary anisotropic fields. Secondly, using a vector field to control 
how $\mathbf{H}$ varies means that the resulting correlation structure can be 
partially visualized from the vector field. Thirdly, when $\gamma >0$ this 
construction guarantees that $\mathbf{H}$ is everywhere positive definite.

\section{Inference}
\label{sec:Inference}
This section begins with a discussion of the parametrization of the model
and the derivation of the posterior distribution. Then the properties
of the inference scheme are discussed through some examples with simulated data.

\subsection{Posterior distribution and parametrization}
The first step for inference is to introduce parameters that control the 
behaviour of the coefficients in Equation~\eqref{eq:fullSPDE} and in turn
the behaviour of the GMRF. The way this is done is by expanding each of
the functions in a basis and use a linear combination of the basis functions
weighted by parameters. For \(\kappa^2\) only one parameter, say
\(\theta_1\), is needed as it is assumed constant, but for the function
\(\mathbf{H}\) a vector of parameters \(\boldsymbol{\theta}_2\) is needed.
Set \(\boldsymbol{\theta} = (\theta_1, \boldsymbol{\theta}_2^\mathrm{T})\)
and give it a prior \(\boldsymbol{\theta}\sim \pi(\boldsymbol{\theta})\).
Then for each value of \(\boldsymbol{\theta}\), the discretization
in Appendix~\ref{sec:discScheme} is used to construct the GMRF
\(\boldsymbol{u}|\boldsymbol{\theta} \sim \mathcal{N}(\boldsymbol{0}, \mathbf{Q}(\boldsymbol{\theta})^{-1})\). 
Combine the prior of $\boldsymbol{\theta}$ with this conditional distribution
to find the joint distribution of the parameters and $\boldsymbol{u}$.
Together with a model for how an observation \(\mathbf{y}\) is made from the 
underlying GMRF this forms a hierarchical spatial model. The relationship
between \(\boldsymbol{y}\) and \(\boldsymbol{u}\) is chosen to be particularly
simple, namely that linear combinations of \(\mathbf{u}\) are observed
with Gaussian noise,
\[
\boldsymbol{y}|\boldsymbol{u} \sim \mathcal{N}(\mathbf{A}\boldsymbol{u},\mathbf{Q}_\mathrm{N}^{-1}),
\]
where $\mathbf{Q}_\mathrm{N}$ is a known precision matrix.

The purpose of the hierarchical model is to do inference
on $\boldsymbol{\theta}$ based on an observation of $\boldsymbol{y}$. With a
Gaussian latent model it is possible to integrate out the
latent field \(\boldsymbol{u}\) exactly and this leads to the log-posterior
\begin{eqnarray}
	\lefteqn{\log(\pi(\boldsymbol{\theta}|\boldsymbol{y})) =} \notag\\
	& & \mathrm{Const}+\log(\pi(\boldsymbol{\theta}))+\frac{1}{2}\log(|\mathbf{Q}(\boldsymbol{\theta})|) \notag \\
	& & { } -\frac{1}{2}\log(|\mathbf{Q}_\mathrm{C}(\boldsymbol{\theta})|)+\frac{1}{2}\boldsymbol{\mu}_C(\boldsymbol{\theta})^\mathrm{T}\mathbf{Q}_\mathrm{C}(\boldsymbol{\theta})\boldsymbol{\mu}_C(\boldsymbol{\theta}),
	\label{eq:postty}
\end{eqnarray}
where $\mathbf{Q}_\mathrm{C}(\boldsymbol{\theta}) = \mathbf{Q}(\boldsymbol{\theta})+\mathbf{A}^\mathrm{T}\mathbf{Q}_\mathrm{N} \mathbf{A}$ and $\boldsymbol{\mu}_\mathrm{C}(\boldsymbol{\theta}) = \mathbf{Q}_\mathrm{C}(\boldsymbol{\theta})^{-1}\mathbf{A}^\mathrm{T}\mathbf{Q}_\mathrm{N} \boldsymbol{y}$.
From the above expression one can see that the posterior distribution of 
$\boldsymbol{\theta}$ contains terms which are hard to handle analytically. 
It is hard to say anything about both the determinants and the quadratic term 
as functions of $\boldsymbol{\theta}$. Therefore, the inference is done 
numerically. The model is on a form which could be handled by the INLA
methodology~\cite{Rue2009}, but at the time of writing the R-INLA 
software\footnote{\url{www.r-inla.org}} does not have the model implemented. Instead
the parameters are estimated  with maximum a posteriori estimates
based on the posterior density given in Equation~\eqref{eq:postty}. In
addition, the standard deviations are estimated  from the square roots
of the diagonal elements of the observed information matrix.
 
The parametrization of \(\mathbf{H}\) introduced in the previous section employs
a pre-defined vector field and a parameter \(\beta\) that controls the magnitude
of the anisotropy due to this vector field. This is a useful representation for
achieving a desired dependence structure, but in a inference setting there
may not be any pre-defined vector field to input. Therefore, the vector field
itself must be estimated. In this context the decomposition of \(\mathbf{H}\) 
introduced in Section~\ref{sec:Preliminaries},
\[
	\mathbf{H}(\boldsymbol{s}) = \gamma \mathbf{I}_2 + \boldsymbol{v}(\boldsymbol{s})\boldsymbol{v}(\boldsymbol{s})^\mathrm{T},
\]
is more useful. For inference it is necessary to control the vector
field by a finite number of parameters. The simple case of a
constant matrix requires 3
parameters. Use parameters \(\gamma\), \(v_1\) and \(v_2\) and write
\[
	\mathbf{H}(\boldsymbol{s}) \equiv \gamma \mathbf{I}_2 + \begin{bmatrix} v_1 \\ v_2 \end{bmatrix}\begin{bmatrix}v_1 & v_2 \end{bmatrix}.
\]

If $\mathbf{H}$ is not constant, it is necessary to parametrize the 
vector field $\boldsymbol{v}$ in some manner. Any vector field is 
possible for $\boldsymbol{v}$, so a basis which can generate any vector 
field is desirable. The Fourier basis possesses this
property, but is only one of many possible choices.
Let the domain be $[0,A]\times[0,B]$ and assume that $\boldsymbol{v}$ is a differentiable, periodic
vector field on the domain. Then each component of the vector field can be written as a Fourier
series of the form
\[
	\sum_{(k,l)\in \mathbb{Z}^2}C_{k,l}\exp\left[2\pi i\left(\frac{k}{A}x+\frac{l}{B}y\right)\right],
\]
where $i$ is the imaginary unit. But since the components are real-valued, 
each of them can also be written as a real $2$-dimensional Fourier series of the form
\[
A_{0,0}+\sum_{(k,l)\in E} \left[A_{k,l}\cos\left[2\pi\left(\frac{k}{A}x+\frac{l}{B}y\right)\right] + B_{k,l} \sin\left[2\pi\left(\frac{k}{A}x+\frac{l}{B}y\right)\right]\right],
\]
where the set $E\subset\mathbb{Z}^2$ is given by
\[
	E = (\mathbb{N}\times\mathbb{Z})\cup (\{0\}\times\mathbb{N}).
\] 

Putting these Fourier series together gives 
\begin{eqnarray}
	\lefteqn{\boldsymbol{v}(\boldsymbol{s}) =} \notag \\
	& & \begin{bmatrix}A_{0,0}^{(1)} \\ A_{0,0}^{(2)} \end{bmatrix}+\sum_{(k,l)\in E} \begin{bmatrix} A_{k,l}^{(1)} \\ A_{k,l}^{(2)} \end{bmatrix}\cos\left[2\pi\left(\frac{k}{A}x+\frac{l}{B}y\right)\right] + \notag\\
	& &	\sum_{(k,l)\in E} \begin{bmatrix} B_{k,l}^{(1)} \\ B_{k,l}^{(2)}\end{bmatrix} \sin\left[2\pi\left(\frac{k}{A}x+\frac{l}{B}y\right)\right],
	\label{chap3:eq:fullFour}
\end{eqnarray}
where $A_{k,l}^{(1)}$ and $B_{k,l}^{(1)}$ are the coefficients for the first component of $\boldsymbol{v}$ and 
$A_{k,l}^{(2)}$ and $B_{k,l}^{(2)}$ are the coefficients of the second component. 
This gives 2 coefficients when only the zero-frequency is included, then 18 parameters
when the $(0,1)$, $(1,-1)$, $(1,0)$ and $(1,1)$ frequencies are included. When
the number of frequencies used in each direction doubles, the number
of required parameters quadruples.

\subsection{Inference on simulated data}
In this section we consider data generated from a known set of parameters. 
The prior used is an improper prior that disallows illegal parameter values. 
It is uniform on \((0,\infty)\) for \(\gamma\) and uniform on
\(\mathbb{R}\) for the rest of the parameters in \(\mathbf{H}\).
The first issue to investigate is whether it is possible to estimate the
stationary model with exactly observed data and whether the approximate
estimation scheme performs well.

\begin{exmp}
\label{exmp:aniInf}
	Use the SPDE
	\begin{equation}
		\label{eq:aniInf}
		u(\boldsymbol{s}) - \nabla\cdot\mathbf{H}\nabla u(\boldsymbol{s}) = \mathcal{W}(\boldsymbol{s}), \qquad \boldsymbol{s}\in[0,20]\times[0,20],
	\end{equation}
	where $\mathcal{W}$ is a standard Gaussian white noise process and
	$\mathbf{H}$ is a $2 \times 2$ matrix, with periodic boundary conditions. Let
	\[
		\mathbf{H} = 3 \mathbf{I}_2 + 2\boldsymbol{v}\boldsymbol{v}^\mathrm{T},
	\]
	with $\boldsymbol{v} = (1,\sqrt{3})/2$. This means that $\mathbf{H}$ has eigenvector
	$\boldsymbol{v}$ with eigenvalue $5$ and an eigenvector orthogonal to $\boldsymbol{v}$ with
	eigenvalue $3$. Construct the GMRF on a $100 \times 100$ grid.

	One observation of the solution is shown in Figure~\ref{fig:aniInf}. Assume
	that the fact that $\mathbf{H}$ is constant is known, but that its value is not. Then using
	the decomposition from the previous sections one can write
	\[
		\mathbf{H} = \gamma \mathbf{I}_2 + \begin{bmatrix}v_1 \\ v_2\end{bmatrix}\begin{bmatrix}v_1 & v_2 \end{bmatrix},
	\]
	where $\gamma$, $v_1$ and $v_2$ are the parameters. Since the process is assumed to
	be exactly observed, we can use the distribution of
	\(\boldsymbol{\theta}|\boldsymbol{u}\). This gives the posterior estimates shown in 
	Table~\ref{tab:aniInfConst}. From the table one can see
	that all the estimates are accurate to one digit, and within one standard deviation of 
	the true value. Actually, this decomposition of $\mathbf{H}$ is 
	invariant to changing $\boldsymbol{v}$ with $-\boldsymbol{v}$, so
	there are two choices of parameters that means the same.

	\begin{figure}
		\centering
		\includegraphics[width=6cm]{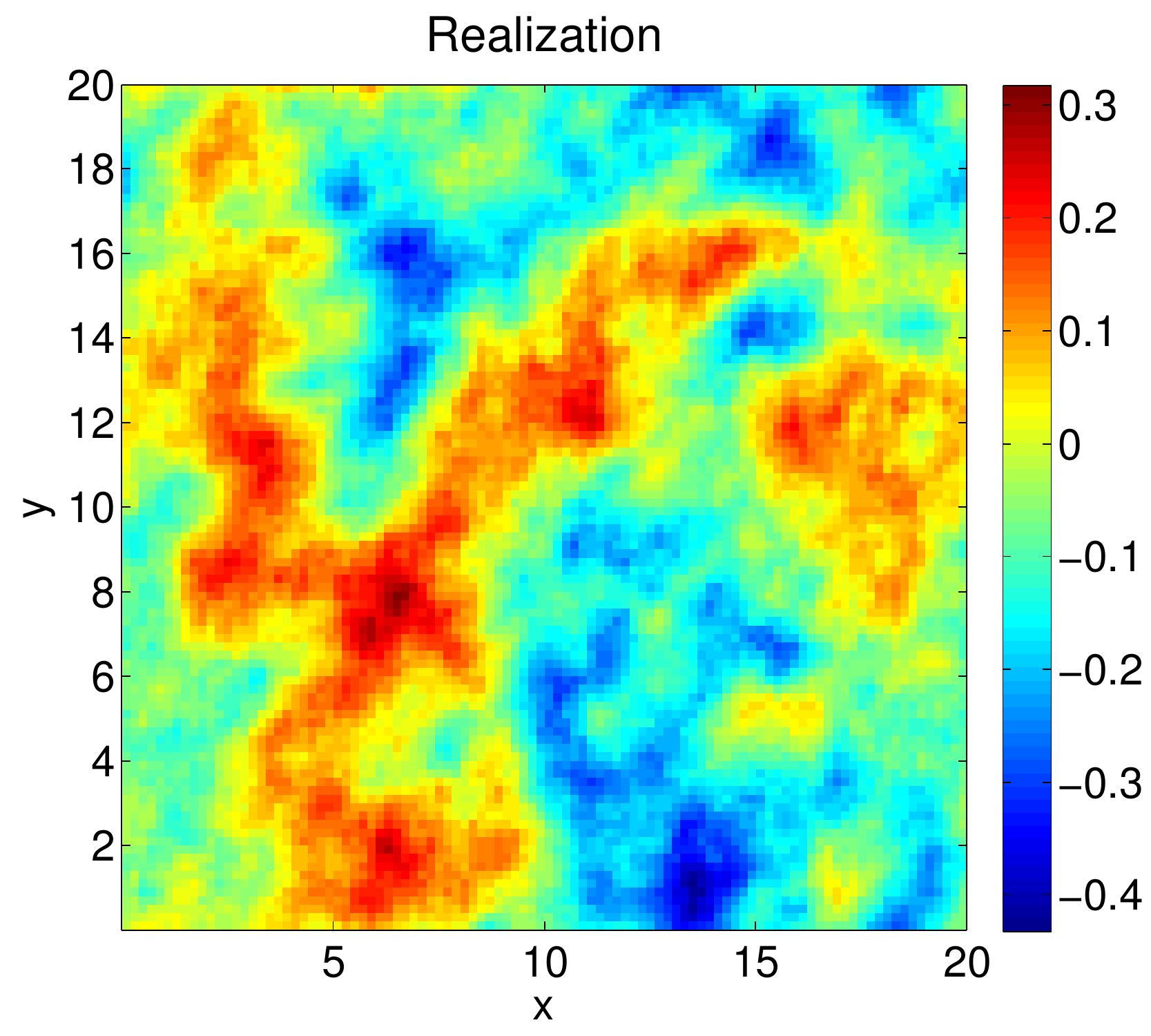}
		\caption{One realization of the solution of the SPDE in Example~\ref{exmp:aniInf}.}
		\label{fig:aniInf}
	\end{figure}

	\begin{table}
		\centering
		\caption{\sf Parameter estimates for Example~\ref{exmp:aniInf}.}
		\begin{tabular}{llll}
			\textbf{Parameter}	& \textbf{True value}	& \textbf{Estimate}	& \textbf{Std.dev.}  \\
			$\gamma$		& 3			& 2.965			& 0.070              \\
			$v_1$			& 0.707			& 0.726			& 0.049		     \\
			$v_2$			& 1.225 		& 1.231         	& 0.039		     
		\end{tabular}
		\label{tab:aniInfConst}
	\end{table}

	The biases in the estimates were evaluated by generating 10000 datasets
	from the true model and estimating the parameters for each dataset. The estimated bias was less than or equal to 
	\(0.1\%\) of the true value for each parameter. Additionally, the 
	sample standard deviations based on the estimation of the parameters
	for each of the 
	10000 datasets 	were \(0.070\), \(0.050\) and \(0.039\) for \(\gamma\), \(v_1\) and \(v_2\), respectively. Each one corresponds well to the 
	corresponding approximate standard deviation, computed 
	via the observed
	information matrix as described in the previous section, that is shown
	in Table~\ref{tab:aniInfConst}.
\end{exmp}

In the above example it is possible to estimate the model, 
but this is under the assumption that
it is known beforehand that the model is stationary. In general, it is
not reasonable to be able to know this beforehand. Therefore, the estimation
is repeated for the more complex model developed in the previous sections
which allows for significant non-stationarity controlled through a vector field. 
The intention is to evaluate whether the more complex model
is able to detect that the true model is a stationary model and if there 
are identifiability issues.

\begin{exmp}
	\label{exmp:15par}
	Use the same SPDE and observation as in Example~\ref{exmp:aniInf}, but
	assume that it is not known that $\mathbf{H}$ is constant. Add the 
	terms in the Fourier series corresponding to the next 
	frequencies, $(k,l) = (0,1)$, $(k,l)=(1,-1)$, $(k,l) = (1,0)$ 
	and $(k,l)=(1,1)$. The observation is still assumed to be exact, but
	there are 16 additional parameters, 4 additional 
	parameters for each frequency.

	First two arbitrary starting positions are chosen for the optimization. 
	The first is $\gamma = 3.0$ and all other parameters at $0.1$. And the 
	second is $\gamma = 3.0$, $A_{0,0}^{(1)} = 0.1$, $A_{0,0}^{(2)} = 0.1$ and 
	all other parameters equal to $0$. For both of these starting points the 
	optimization converges to non-global maximums. Parameter estimates and
	approximate standard deviations are not show, but Figure~\ref{fig:aniInf-vec}
	shows the two different vector fields found.

	A third optimization is done with starting values close to the correct
	parameter values. This gives a vector field close to the actual one, with
	estimates for  $\gamma$, $A_{0,0}^{(1)}$ and $A_{0,0}^{(2)}$ that agree with
	the ones in Example~\ref{exmp:aniInf} to two digits. The other frequencies 
	all had coefficients close to zero, with the largest having an absolute value
	of $0.058$.

	\begin{figure}
		\centering
		\subfigure[Wrong maximum.]{
			\includegraphics[width=5cm]{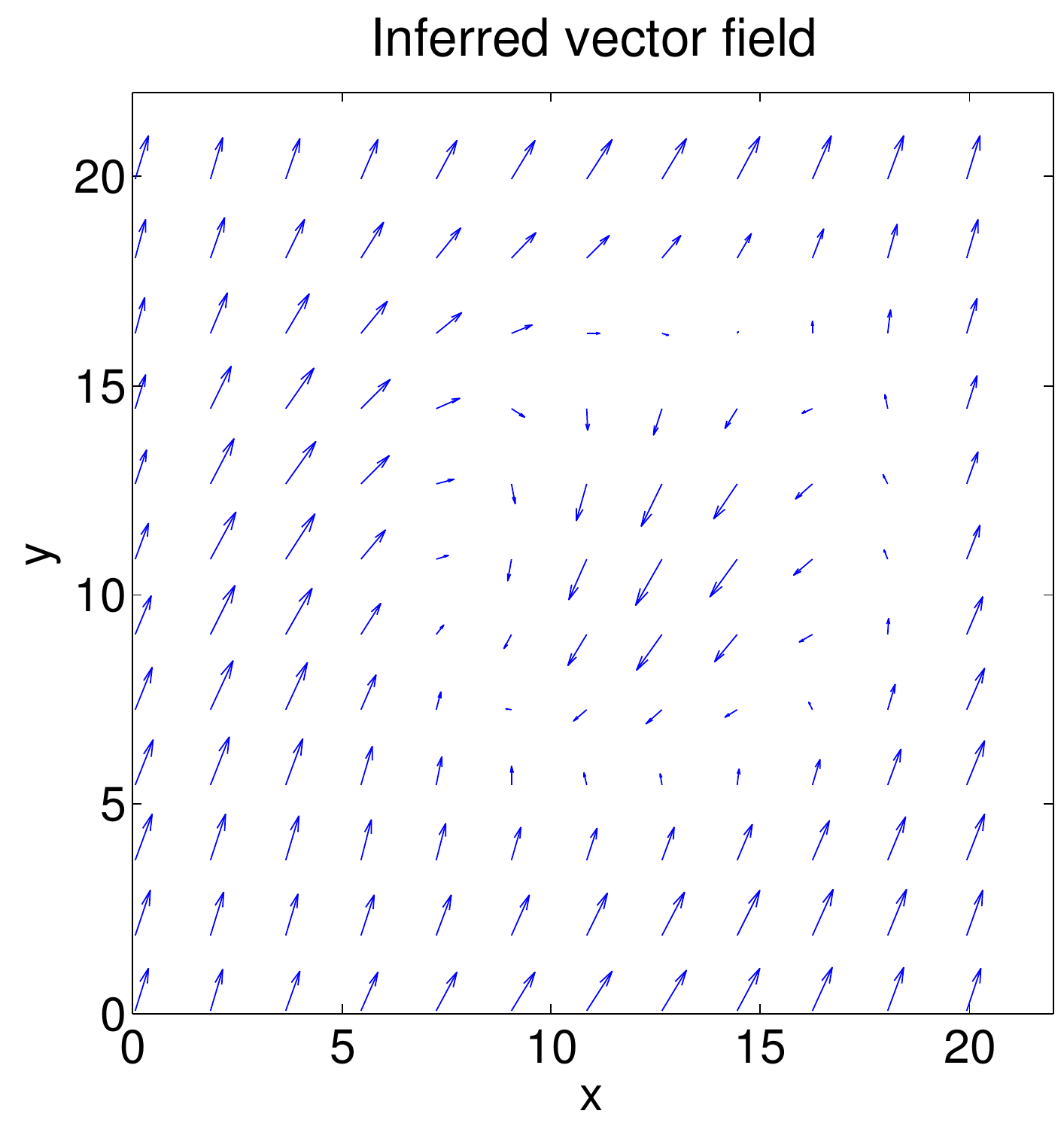}
			\label{fig:aniInf-vec1}
		} 
		\subfigure[Wrong maximum.]{
			\includegraphics[width=5cm]{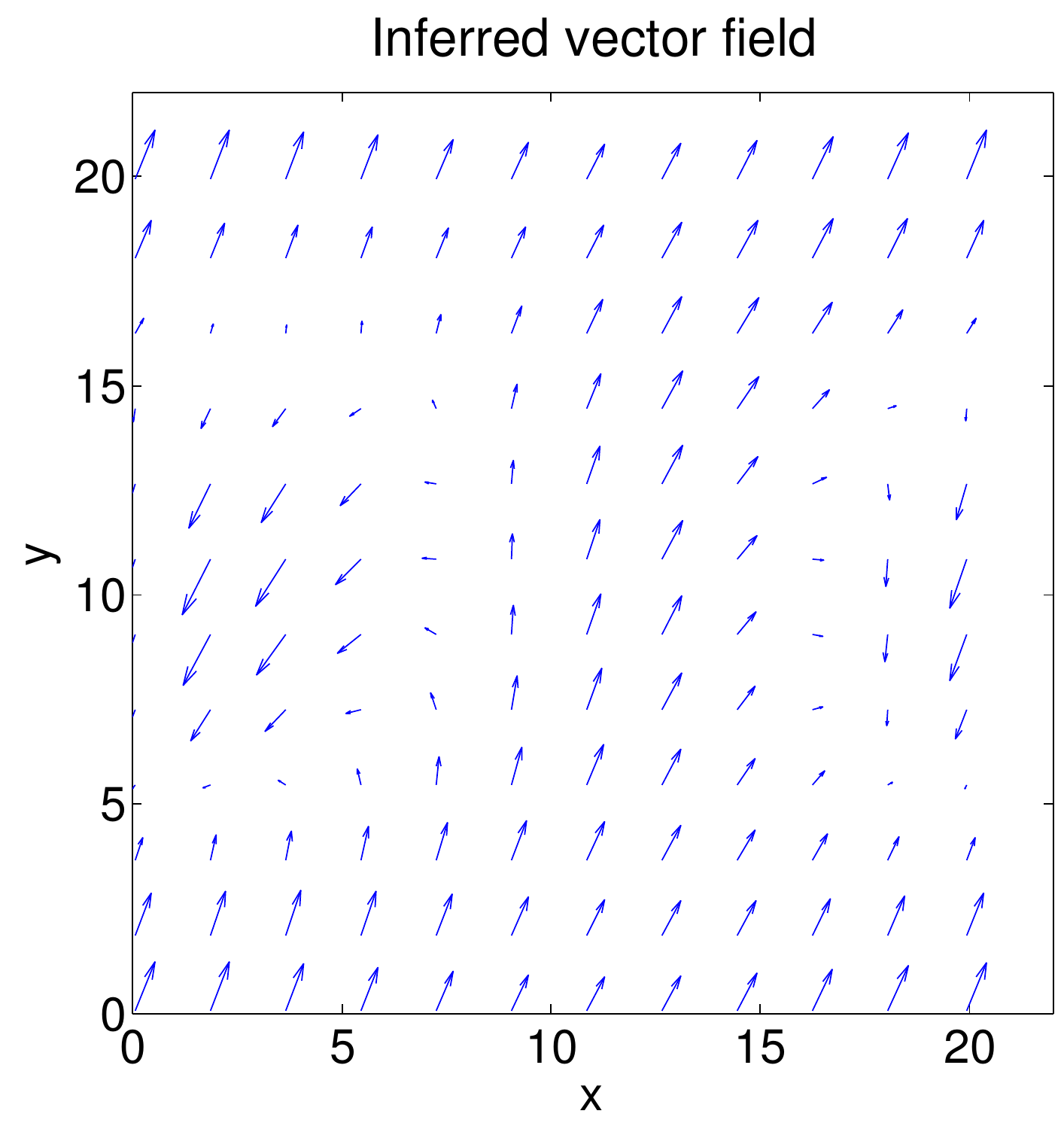}
			\label{fig:aniInf-vec2}
		}
		\caption{Two different local maxima found for the vector field. The vector field in \subref{fig:aniInf-vec1}
			 has a lower value for the posterior distribution of the parameters than the vector field in \subref{fig:aniInf-vec2} and
			 both has lower value than the actual maximum.}
		\label{fig:aniInf-vec}
	\end{figure}

	The results illustrate a difficulty with estimation caused by the
	the inherent non-identifiability of the sign of the vector field.
	The true vector field is constant and in Figure~\ref{fig:aniInf-vec} 
	one can see that each vector field has large parts
	which has the correct appearance if one only considers the lines 
	defined by the arrows and not in which of the two possible directions
	that the arrow points. The positions where the vector field
	is wrong are smaller areas where the vector field flips its direction.
	The problem is that it is difficult to reverse this flipping as it
	requires moving through states with smaller likelihood. Thus creating
	undesirable local maximums. One approach to improving the 
	situation would be to force an apriori preference for vector 
	fields without abrupt changes. That is to introduce a prior 
	which forces higher frequencies
	of the Fourier basis to be less desirable. This is an issue that
	needs to be addressed for an application and is briefly discussed
	in Section~\ref{sec:Extensions}.

	By acknowledging the issue and starting close to the true value, one
	can do repeated simulations of datasets and prediction of parameters
	to evaluate how well the non-stationary model captures the fact 
	that the true model is 
	stationary and see if there is any consistent bias. 1000 datasets
	were simulated and the estimation of the parameters was done for each
	dataset with a starting value close to the true value. This gives
	the result summarized in the boxplot in 
	Figure~\ref{fig:biasInVectorField}. There does not appear to be
	any significant bias and the parameters that give non-stationarity
	are all close to zero.

	\begin{figure}
		\centering
		\includegraphics[width=7cm]{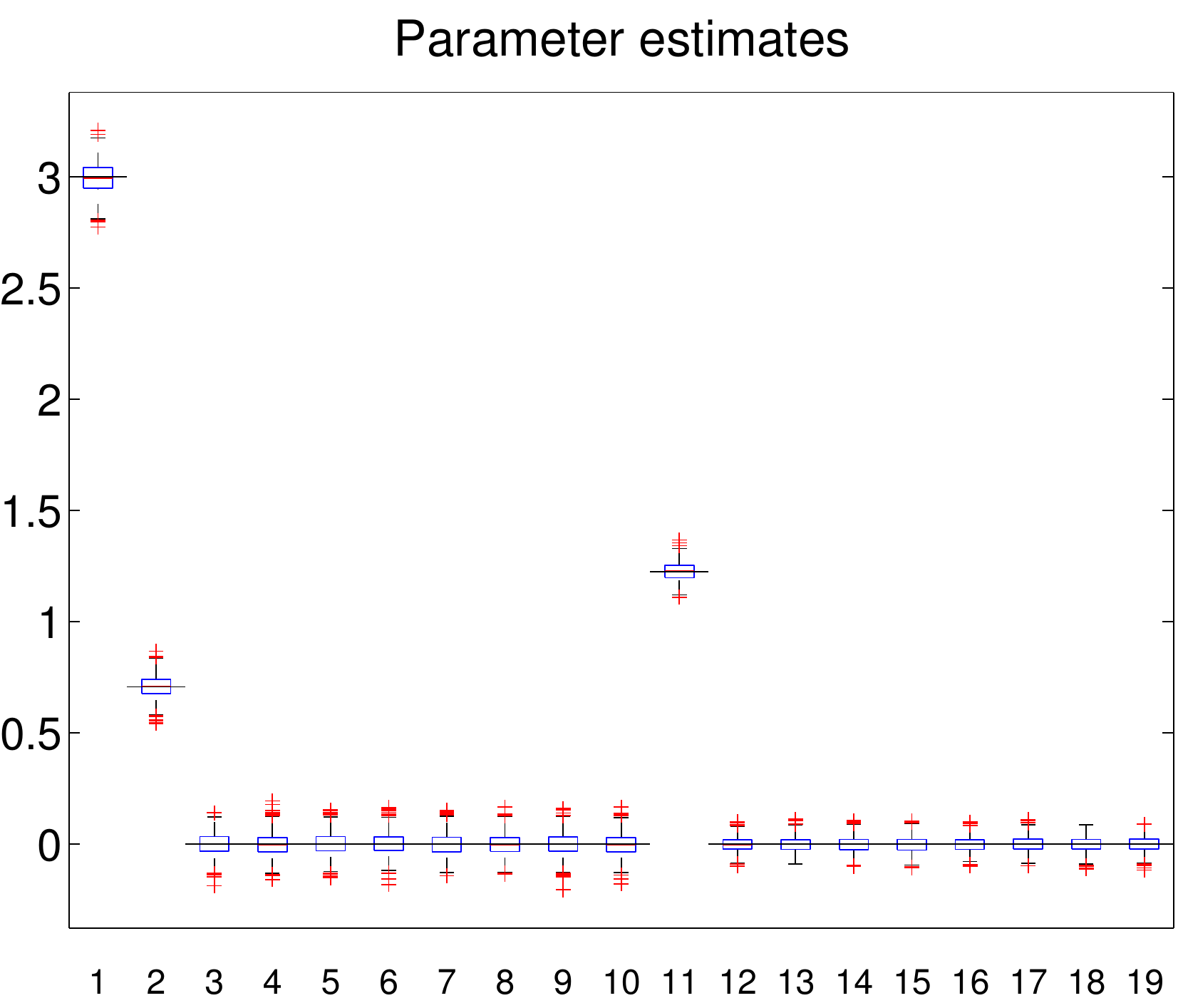}
		\caption{Boxplot of estimated parameters for 1000 simulated
				 datasets in Example~\ref{exmp:15par}. 
				 Parameters~1, 2 and 11 corresponds to \(\gamma\),
				 \(v_1\) and \(v_2\), respectively. The red lines
				 inside the boxes specify the medians and the longer
				 black lines through the boxes specify the true parameter
				 values. In most of the boxes the red line is completely
				 covered by the black line.}
		\label{fig:biasInVectorField}
	\end{figure}
\end{exmp}

The example shows that there are issues in estimating the anisotropy in the
non-stationary model due to the non-identifiability of the sign of the 
vector field, but that if one avoids the local maximums the estimated model 
is close to the true stationary model in this case. In addition, 
there is a significant
increase in computation time when increasing the parameter space from
\(3\) to \(19\) parameters. The computation time required is increased 
by a factor of approximately $10$.

A high-dimensional model increases the flexibility, but as seen above also
adds additional difficulties. In situations where there is a physical
explanation of the additional dependence in one direction, it would be 
desirable to do a simpler model with one parameter for the baseline 
isotropic effect and one parameter specifying the degree of anisotropy 
caused by a pre-defined vector field such as in Example~\ref{exmp:nonStat}.
This presents a simplification from the previous inference examples because
the vector field itself does not need to be estimated.

\begin{exmp}
	\label{exmp:finf}
	Use a $100 \times 100$ grid of $[0,20]^2$ and
	periodic boundary conditions for the SPDE in Equation~\eqref{eq:fullSPDE}.
	Let $\kappa^2$ be equal to $1$ and let $\mathbf{H}$ be parametrized
	as
	\[
		\mathbf{H}(\boldsymbol{s}) = \gamma \mathbf{I}_2 + \beta \boldsymbol{v}(\boldsymbol{s})\boldsymbol{v}(\boldsymbol{s})^\mathrm{T},
	\]
	where $\boldsymbol{v}$ is the vector field from Example~\ref{exmp:nonStat}.

	Figure~\ref{fig:inf-fobs} shows one observation of the solution with
	$\gamma = 0.5$ and $\beta = 5$. In this case one expects that it is
	possible to make accurate estimates about $\gamma$ and $\beta$ as the
	situation is simpler than in the previous example.

	\begin{figure}
		\centering
		\includegraphics[width=6cm]{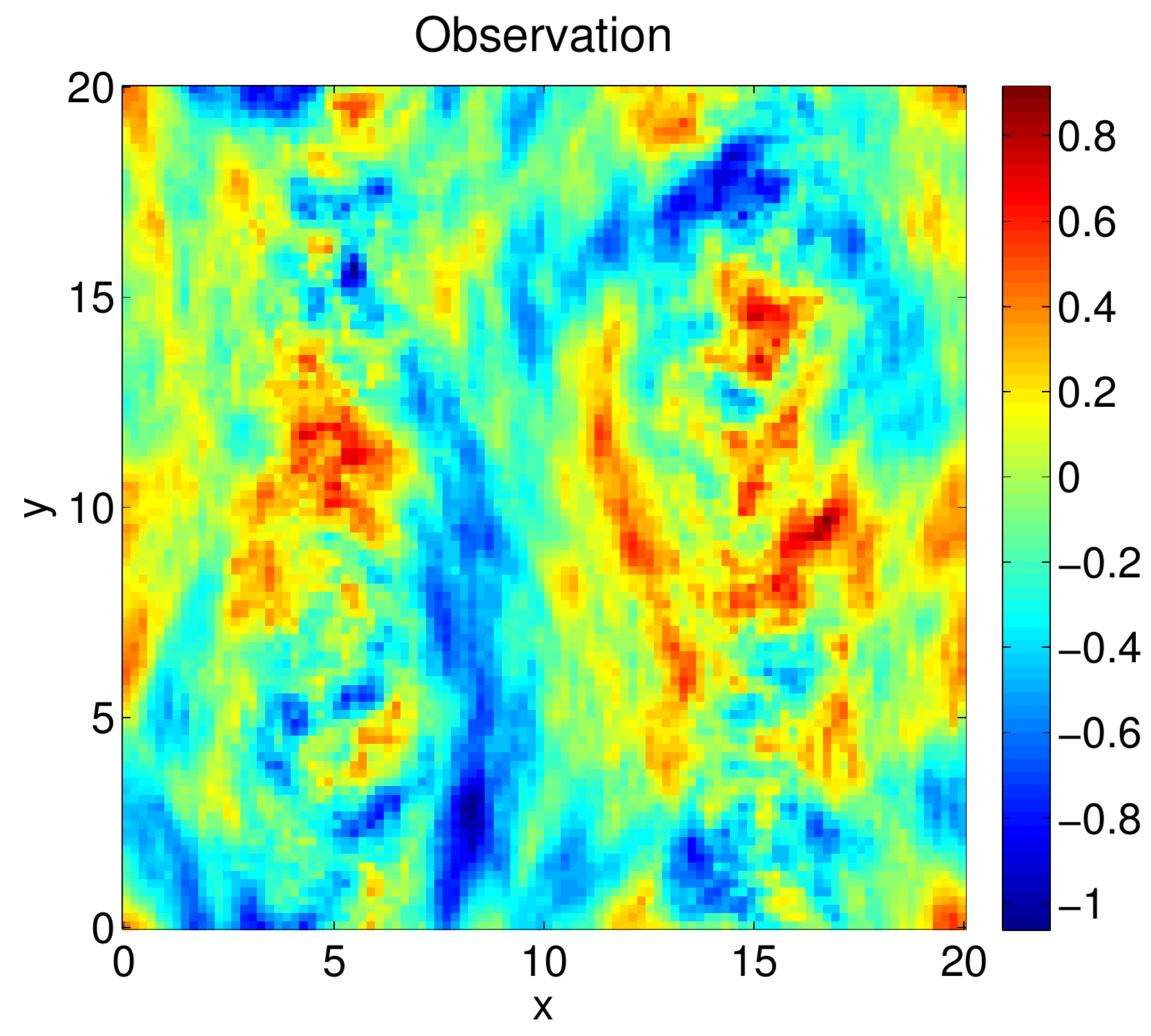}
		\caption{An observation of the SPDE in Equation~\eqref{eq:fullSPDE} on a $100 \times 100$ regular 
		grid of $[0,20]^2$ with periodic boundary conditions, $\kappa^2 = 1$ and 
		$\mathbf{H}(\boldsymbol{s}) = 0.5{\sf \bf I}_2+5\boldsymbol{v}(\boldsymbol{s})\boldsymbol{v}(\boldsymbol{s})^\mathrm{T}$, 
		where $\boldsymbol{v}$ is the vector field in Example~\ref{exmp:nonStat}.}
		\label{fig:inf-fobs}
	\end{figure}

	The estimated parameters are shown in Table~\ref{tab:finf}. 
	From the table one can see that the estimates for both $\gamma$ and $\beta$
	are quite accurate, which is reflected both in the actual value of the estimates and the approximated standard deviations. The estimates for both 
	$\gamma$ and $\beta$ are accurate to $2$ digits. In a similar way as
	in the previous example, the bias is estimated to be less than \(0.02\%\) 
	for each each parameter, and the sample standard deviation from 
	estimation over many datasets is 0.008 and 0.08 for \(\gamma\) and
	\(\beta\) respectively.

	\begin{table}
		\centering
		\caption{\sf Posterior inference on parameters in Example~\ref{exmp:finf}.}
		\begin{tabular}{llll}
			{\bf Parameter}	&	{\bf True value}	& {\bf Estimate}	& {\bf Std.dev.} \\
			$\gamma$	&	0.5			& 0.5012		& 0.0081 \\
			$\beta$		&	5			& 5.014			& 0.084
		\end{tabular}
		\label{tab:finf}
	\end{table}
\end{exmp}

The above example does not have the same issues as Example~\ref{exmp:15par} where
the vector field itself must be estimated. The example shows that when using 
only the $\gamma \mathbf{I}_2$ term and fixed vector field where only the
magnitude of the effect is controlled by a parameter \(\beta\), the estimates 
of the parameters are quite accurate. The accuracy of the estimates 
will of course depend on the vector field used.

In a more realistic situation the actual basis needed for the vector field 
is not known and there is observation noise. In the following example the 
estimation is compared when all required frequencies are included and 
when only a subset of the required frequencies of the Fourier basis is included.

\begin{exmp}
\label{exmp:noisy-inf}
	Use a $100 \times 100$ grid of $[0,20]^2$ and
	periodic boundary conditions for the SPDE in Equation~\eqref{eq:fullSPDE}.
	Let $\kappa^2$ be equal to $1$ and let $\mathbf{H}$ be given
	as
	\[
		\mathbf{H}(\boldsymbol{s}) = \mathbf{I}_2 + \boldsymbol{v}(\boldsymbol{s})\boldsymbol{v}(\boldsymbol{s})^\mathrm{T},
	\]
	where $\boldsymbol{v}$ is the vector field
	\[
		\boldsymbol{v}(x,y) = \begin{bmatrix} 2+\cos\left(\frac{\pi}{10}x\right) \\ 3+2\sin\left(\frac{\pi}{10}y\right)+\sin\left(\frac{\pi}{10}(x+y)\right)\end{bmatrix}.
	\]

	One observation with i.i.d.\ Gaussian noise with precision $400$ is shown
	in Figure~\ref{fig:noisy-inf-obs}.
	Based on this realization it is desired to estimate the correct value of 
	$\gamma$ and the correct vector field
	$\boldsymbol{v}$ in the parametrization
	\[
		\mathbf{H}(\boldsymbol{s}) = \gamma \mathbf{I}_2 + \boldsymbol{v}(\boldsymbol{s})\boldsymbol{v}(\boldsymbol{s})^\mathrm{T}.
	\]
	First use only one extra frequency in each direction, that is only the frequencies $(0,0)$, $(0,1)$ and $(1,0)$.
	This gives the estimated vector field shown in Figure~\ref{fig:noisy-vec-some}. Then add the missing frequency
	and use the frequencies $(0,0)$, $(0,1)$, $(1,0)$ and $(1,1)$. This gives the estimated vector field shown in
	Figure~\ref{fig:noisy-vec-all}. The true vector field is shown in Figure~\ref{fig:noisy-vec-real}.

	\begin{figure}
		\centering
		\includegraphics[width=6cm]{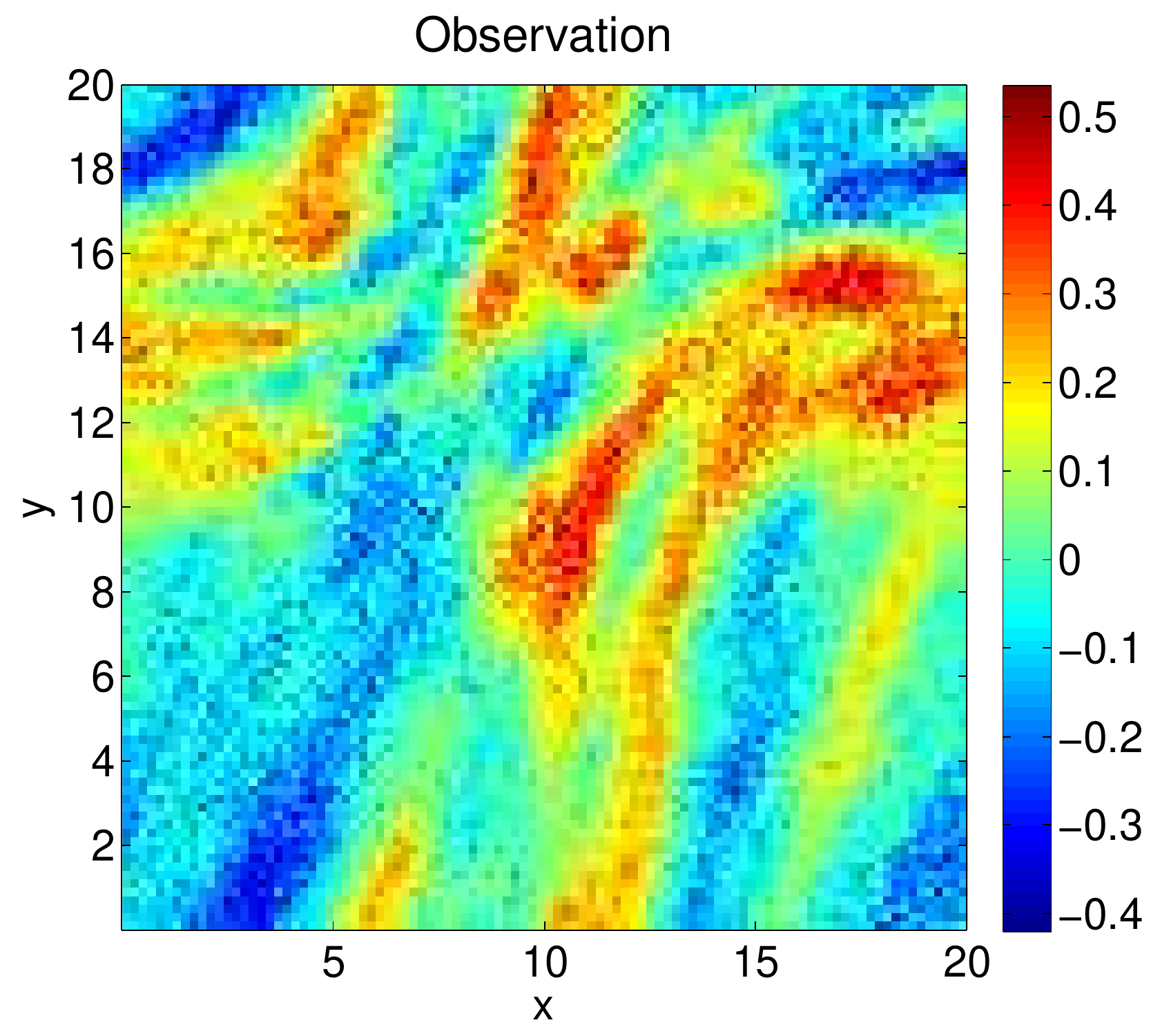}
		\caption{An observation of the SPDE in Example~\ref{exmp:noisy-inf}
			 with i.i.d.\ Gaussian white noise with precision $400$.}
		\label{fig:noisy-inf-obs}
	\end{figure}

	Both estimated vector fields are quite similar to the true vector field, and 
	the $\gamma$ parameter 	was estimated to $1.14$ in the first case and $1.09$ in the latter case. There is a clear bias in the estimate of \(\gamma\), but
	this must be expected as there is a need to compensate for the lacking 
	frequencies. All parameter values were estimated, but are not shown. 
	For the first case many parameters is more than two standard deviations 
	from their correct values and in the second case this only 
	happens for one parameter. For each case the difference between the 
	true \(\mathbf{H}\) and the estimated \(\hat{\mathbf{H}}\) is calculated
	through
	\[
		\frac{1}{100}\sqrt{\sum_{i=1}^{100}\sum_{j=1}^{100} \left\lvert\left\lvert \mathbf{H}(\boldsymbol{s}_{i,j})-\hat{\mathbf{H}}(\boldsymbol{s}_{i,j})\right\rvert\right\rvert_2^2},
	\]
	where \(\boldsymbol{s}_{i,j}\) are the centres of the cells in the
	grid and \(||\cdot||_2\) denotes the 2-norm. 
	The case with frequencies \((0,0)\), \((0,1)\) and \((1,0)\) gives
	7.9 and the case with frequencies \((0,0)\), \((0,1)\), \((1,0)\) and
	\((1,1)\) gives 1.5.

	\begin{figure}
		\centering
		\subfigure[$(0,0)$, $(0,1)$ and $(1,0)$ frequencies]{
			\includegraphics[width=6cm]{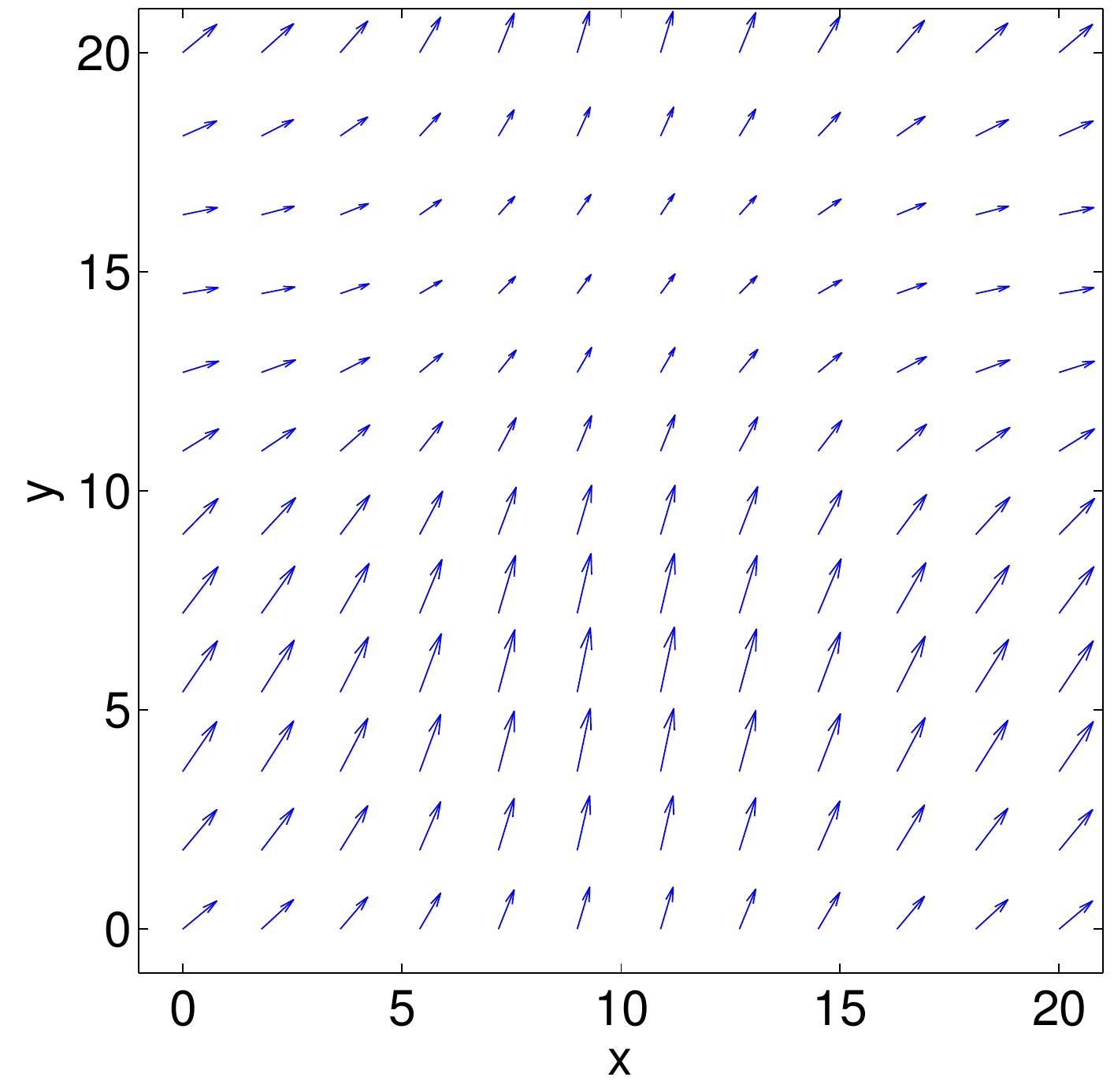}
			\label{fig:noisy-vec-some}
		}
		\subfigure[$(0,0)$, $(0,1)$, $(1,0)$ and $(1,1)$ frequencies]{
			\includegraphics[width=6cm]{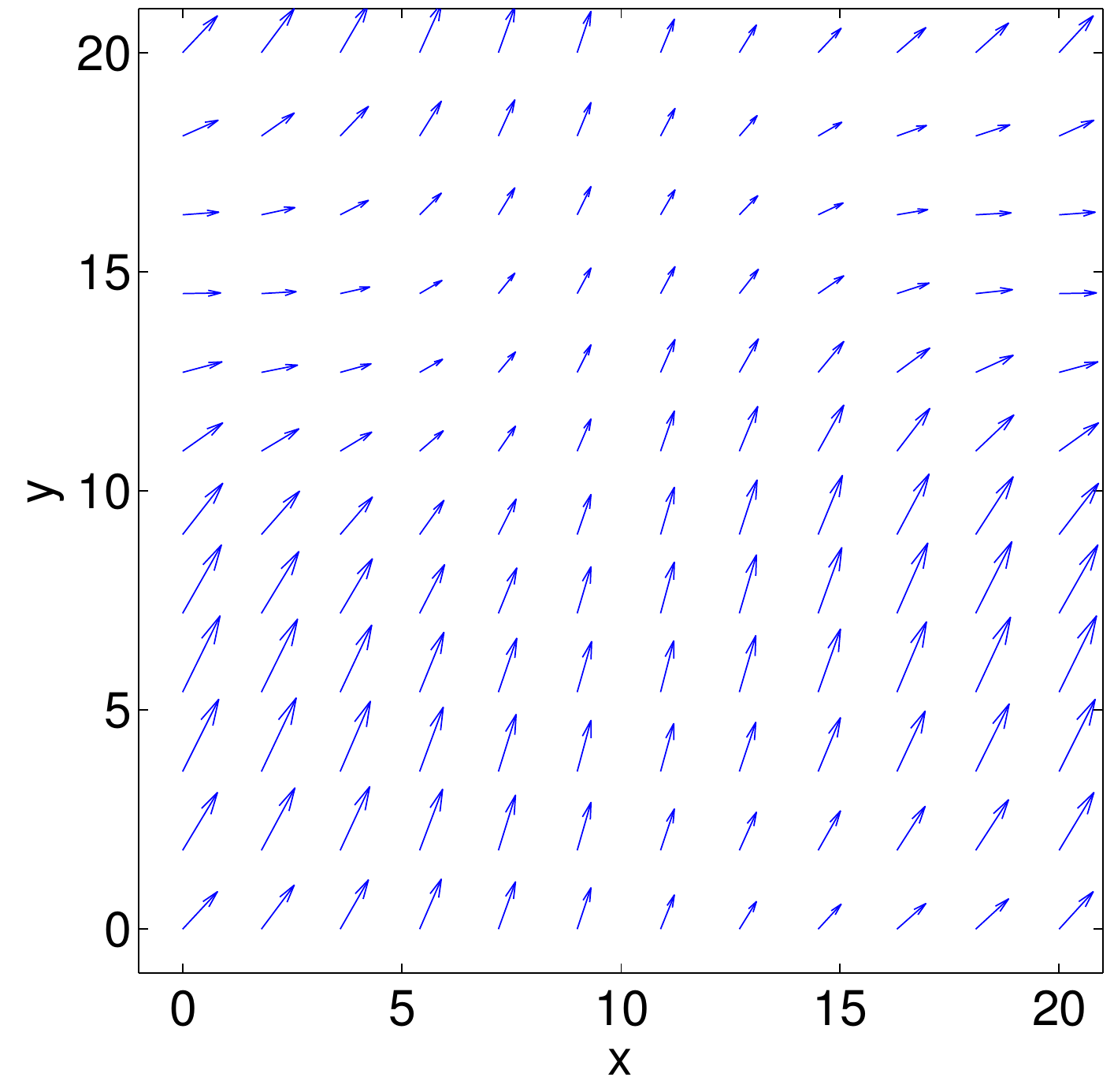}
			\label{fig:noisy-vec-all}
		}\\
		\subfigure[The true vector field]{
			\includegraphics[width=6cm]{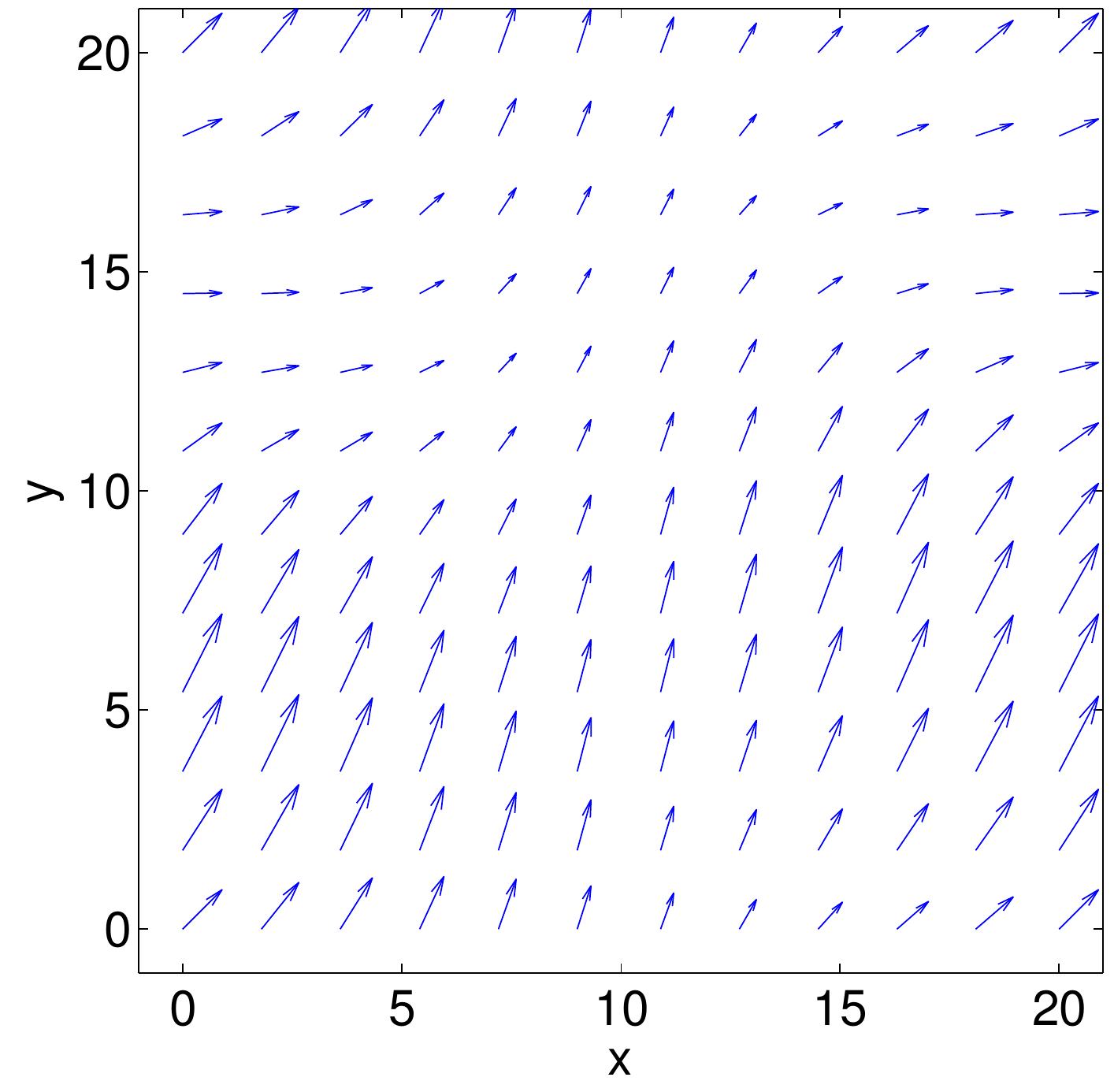}
			\label{fig:noisy-vec-real}
		}
		\caption{True vector field and inferred vector fields in 
			     Example~\ref{exmp:noisy-inf}. Each of the vector fields
			     is scaled with a factor \(0.3\).}
		\label{fig:noisy-vec}
	\end{figure}
\end{exmp}

These examples focus on simple cases where specific issues can be highlighted.
The inherent challenges in estimating a spatially varying direction and strength
are equally important in the more general setting where also 
\(\kappa\) and the baseline effect \(\gamma\) is allowed to vary. The estimation
of the vector field presents an important component that must be dealt with
in any inference strategy for the more general case. 

\section{Extensions}
\label{sec:Extensions}
The class of models discussed in the previous sections offers a
flexible way to introduce and control directional dependence at each
location using a vector field. This is an important step towards a full 
flexible non-stationary 
model for practical applications, but still leaves something to be desired. 
To make the model applicable to real-world datasets it is necessary to also 
make the parameters \(\kappa\) and \(\gamma\) spatially varying functions. 
This results in some control also over the marginal variance and the strength 
of the local baseline component of the anisotropy at each location. A varying
\(\kappa\) is discussed briefly in Section~3.2 in~\ocite{Lindgren2011}.

However, this comes at the cost of two more functions that must be 
inferred together with the vector field \(\boldsymbol{v}\). Which in turn means 
two more functions that need to be expanded into bases. This could be done 
in a similar way as for the vector field with a Fourier basis, but the Fourier
basis does not constitute the only possible choice, and any basis which 
respects the boundary condition could in principle be used. But the amount 
of freedom available by having four spatially varying functions comes at a 
price, and it would be necessary to introduce some apriori restrictions on 
the behaviour of the functions.

In Example~\ref{exmp:15par} the challenge with the non-identifiability
of the sign of the vector field is demonstrated. It would be possible to make 
the situation less problematic by enforcing more structure in the 
estimated vector field. For example, through spline penalties which adds a
preference for components without abrupt changes. Such apriori restrictions 
make sense both from a modelling perspective, in the sense that the properties
should not change to quickly, and from a computational perspective, in
the sense that it is desirable to avoid situations as the one encountered
in the previous section where the direction of the vector field flips.

The full model could be used in a real-world application through a 
three step approach. First, choose an appropriate basis to use for each
function and select an appropriate prior. This means deciding how many 
basis elements one is willing to use, from a computational point of view,
 and how strong the apriori penalties 
needs to be. Second, find the maximum aposteriori estimate of the 
functions \(\kappa\), \(\gamma\), \(v_1\) and \(v_2\). Third, assume the
maximum aposteriori estimates are the true functions and calculate the 
predicted values and prediction variances. The full details of such
an approach is beyond the scope of this paper and is being studied
in current work on an application to annual rainfall data in the conterminous 
US~\cite{Fuglstad2013b}.

Another way forward deals with the interactions of
the functions \(\kappa\), \(\gamma\), \(v_1\) and \(v_2\).
The functions interact in difficult ways to control marginal
variance and to control anisotropy. As seen in Example~\ref{exmp:nonStat} the 
vector field that controls the anisotropic behaviour is also linked to the 
marginal variances of the field. It would be desirable to try to separate 
the functions that are allowed to affect the marginal variances and the 
functions that are allowed to affect the correlation structure. This
may present a useful feature in applications, both for interpretability 
and for constructing priors.

One promising way to greatly reduce this interaction is to extend on the 
ideas presented in Section~3.4 in~\ocite{Lindgren2011}. The section links 
the use of an anisotropic Laplacian to the deformation method 
of~\ocite{Sampson1992}. The link presented is in itself
too restrictive, but the last comments about the connection to metric tensors 
leads to a useful way to rewrite the SPDE in Equation~\eqref{eq:fullSPDE}. 
This is work in progress and involves interpreting the simple SPDE 
\begin{equation}
	[1-\Delta]u = \mathcal{W}
	\label{eq:SIMPLE}
\end{equation}
as an SPDE on a Riemannian manifold with an inverse metric tensor 
defined through the strength of dependence in different directions in a similar way 
as the spatially varying matrix \(\mathbf{H}\). This leads to a slightly
different SPDE, where a separate function, which does not affect
correlation structure, can be used to control marginal standard deviations. 
However, the separation is not perfect since the varying metric tensor 
gives a curved space and thus affects the marginal variances of the solution
of the above SPDE. But the effect
of the metric tensor on marginal standard deviations appears small, and it
appears to be a promising way forward. 

Another issue which is not addressed in the previous sections is how
to define relevant boundary conditions.  For rectangular domains,
periodic boundary conditions as used here are simple to implement, but
a naive use of such conditions will typically not inappropriate in
practical applications due to the resulting spurious dependence
between physically distant locations.  This problem can be partly
rectified by embedding the region of interest into a larger covering
domain, so that the boundary effects are moved away from the region
that directly influences the likelihood function.  It is also possible
to apply Neumann type boundary conditions similar to the ones used by
\ocite{Lindgren2011}.  These are easier to adapt to more general
domains, but they still require a domain extension in order to remove
the influence of the boundary condition on the likelihood.  A more
theoretically appealing, and computationally potentially less
expensive, solution would be to directly define the behaviour of the
field along the boundary so that the models would contain stationary
fields as a neutral case.  Work is underway to design stochastic
boundary conditions to accomplish this, and some of the solutions show
potential for extension to non-stationary models.


\section{Discussion}
\label{sec:Discussion}
The paper explores different aspects of a new class of non-stationary 
GRFs based on local anisotropy. The benefit of the formulation presented is that
it allows for flexible models with few requirements on the parameters. 
Since the GRF is based on an SPDE, there is no need to worry about how to 
change the discretized model in a consistent manner when the grid is refined.
In other words, one does not need to worry
about how the precision matrix must be changed to give a similar
covariance structure when the number of grid points is increased. This is
one of the more attractive features of the SPDE-based modelling.

The focus of the examples has been the matrix \(\mathbf{H}\) introduced
in the Laplace-operator. The examples show that a variety of different
effects can be achieved by using different types of spatially varying
matrices. Constant matrices of the form \(\gamma \mathbf{I}_2\) give 
isotropic random fields and constant matrices of other forms give anisotropic,
stationary random fields. 
As shown in Section~\ref{sec:Examples} the anisotropic fields have
anisotropic Mat\'{e}rn-like covariance functions, through stretching and rotating
the domain, and can be controlled by four parameters. It is possible to
control the marginal variances, the principal directions and the range
in each of the principal directions. A spatially varying \(\mathbf{H}\) 
gives non-stationary random fields. And by using a vector field to 
specify the strength and direction of extra spatial dependence in 
each location, there is a clear connection between the vector field
and the resulting covariance structure. The covariance structure
can be partially visualized from the vector field.

From the examples in Section~\ref{sec:Inference} one can see that sensible
values for the parameters are estimated both with and without noise, except
for problems with multimodality in Example~\ref{exmp:15par}, which uses
a more flexible construction for the vector field than the other examples. 
Additionally, the examples show no significant biases in the estimate.
The last example presents the most challenging case, where the true model
cannot be represented by the model estimated, and is perhaps closest to a 
real scenario. In the example good results are achieved when estimating 
the vector field with only a subset of the frequencies required to fully 
describe it. 

There are many avenues that are not explored in this paper due to 
the fact that it is a first look into a new type of model. The chief
motivation is to explore the class of models both in the sense of
what can be achieved and associated challenges for inference with the 
model. In this paper it is shown that a vector field constitutes a useful way
to control local anisotropy in the SPDE-model of~\ocite{Lindgren2011}.
What remains for a fully flexible spatial model is to allow 
also $\kappa$ and $\gamma$ to be spatially varying functions. However,
this is a simpler task than the anisotropy component since they do 
not require vector fields. For this more complex model there will
be \(4\) spatially varying functions to estimate and an expansion
of each of these functions into a basis will lead to many parameters. 
This means it is necessary to explore ways of dealing with 
high-dimensional estimation problems. Additionally, it remains to 
investigate appropriate choices of priors for use in applications. 
This question is connected with the discussion in 
Section~\ref{sec:Extensions} on an alternative construction of the
model which separates the functions that are allowed to affect 
marginal variances and the functions that allowed to affect 
correlation structure.

\section*{Acknowledgements}
The authors are grateful to Editors  
and referees for their helpful comments 
which improved the manuscript.

\clearpage
\appendix
\numberwithin{equation}{section}
\numberwithin{table}{section}
\numberwithin{figure}{section}
\section{Derivation of precision matrix}
\label{app:DerivationPrecisionMatrix}
\subsection{Formal equation}
The SPDE is
\begin{equation}
	(\kappa^2(\boldsymbol{s}) -\nabla\cdot\mathbf{H}(\boldsymbol{s}))\nabla u(\boldsymbol{s}) = \mathcal{W}(\boldsymbol{s}), \qquad \boldsymbol{s} \in [0,A]\times [0,B],
	\label{eq:mainSPDE}
\end{equation}
where $A$ and $B$ are strictly positive constants, $\kappa^2$ is a scalar 
function, ${\sf \bf H}$ is a $2\times 2$ matrix-valued function,
$\nabla=\left(\frac{\partial}{\partial x},\frac{\partial}{\partial y}\right)$
and $\mathcal{W}$ is a standard Gaussian white noise process. In addition, 
$\kappa^2$ is assumed to be a continuous, strictly positive function and 
$\mathbf{H}$ is assumed to be a continuously differentiable function which 
gives a positive definite matrix $\mathbf{H}(\boldsymbol{s})$ for each 
$\boldsymbol{s}\in[0,A]\times[0,B]$.

Further, periodic boundary conditions are used,  which means that opposite sides 
of the rectangle $[0,A]\times[0,B]$ are identified. This gives 
additional requirements for $\kappa^2$ and $\mathbf{H}$. The values of 
$\kappa^2$ must agree on opposite edges and the values of $\mathbf{H}$ and 
its first order derivatives must agree on opposite edges.
The periodic boundary conditions are not essential to the methodology presented
in what follows, but were chosen to avoid the issue of appropriate boundary 
conditions.

\subsection{Finite volume methods}
\label{sec:finVolMeth}
In the discretization of the SPDE in Equation~\eqref{eq:mainSPDE} a finite 
volume method is employed. The finite volume methods are useful for 
creating discretizations of conservation laws of the form
\[
	\nabla\cdot \boldsymbol{F}(\boldsymbol{x},t) = f(\boldsymbol{x},t),
\]
where $\nabla\cdot$ is the spatial divergence operator. This equation relates 
the spatial divergence of the flux $\boldsymbol{F}$ and the sink-/source-term $f$. 
The main tool in these methods is the use of the divergence theorem
\begin{equation}
	\int_E \! \nabla\cdot\boldsymbol{F} \, \mathrm{d}V = \oint_{\partial E} \! \boldsymbol{F}\cdot \boldsymbol{ n} \,\mathrm{d}\sigma,
	\label{eq:divThm}
\end{equation}
where $\boldsymbol{n}$ is the outer normal vector of the surface $\partial E$ 
relative to $E$.

The main idea is to divide the domain of the SPDE in 
Equation~\eqref{eq:mainSPDE} into smaller parts and consider the resulting 
``flow'' between the different parts. A lengthy treatment of finite volume 
methods is not given, but a comprehensive treatment of the method for 
deterministic differential equations can be found in~\ocite{eymard2000finite}.

\subsection{Derivation}
\label{sec:discScheme}
\begin{figure}
	\centering
	\begin{tikzpicture}[scale=1.7]
	\tikzstyle{every node}=[]
	\tikzstyle{nodeR} = [rotate=90]
	\draw[line width = 1, ->]  (0,0) -- (0,5);
	\draw[line width = 1, ->]  (0,0) -- (5,0);
	\draw[  ]  (1,0) -- (1,2);
	\draw[  ]  (2,0) -- (2,2);
	\draw[  ]  (0,1) -- (2,1);
	\draw[  ]  (0,2) -- (2,2);
	\draw[  ]  (0,3) -- (2,3);
	\draw[  ]  (0,4) -- (2,4);
	\draw[  ]  (1,3) -- (1,4);
	\draw[  ]  (2,3) -- (2,4);
	\draw[  ]  (3,0) -- (3,2);
	\draw[  ]  (4,0) -- (4,2);
	\draw[  ]  (3,1) -- (4,1);
	\draw[  ]  (3,2) -- (4,2);
	\draw[  ]  (3,3) -- (4,3) -- (4,4) -- (3,4) -- (3,3);
	\draw[line width = 0.8, dashed] (1,2.1) -- (1,2.9);
	\draw[line width = 0.8, dashed] (2.1,1) -- (2.9,1);
	\draw[line width = 0.8, dashed] (3.5,2.1) -- (3.5,2.9);
	\draw[line width = 0.8, dashed] (2.1,3.5) -- (2.9,3.5);
	\draw[line width = 0.8, dashed] (2.1,2.1) -- (2.9,2.9);
	\path (0.5,0.5)    node (v1)  {$E_{0,0}$};
	\path (1.5,0.5)    node (v1)  {$E_{1,0}$};
	\path (0.5,1.5)    node (v1)  {$E_{0,1}$};
	\path (1.5,1.5)    node (v1)  {$E_{1,1}$};
	\path (3.5,0.5)    node (v1)  {$E_{M-1,0}$};
	\path (3.5,1.5)    node (v1)  {$E_{M-1,1}$};
	\path (0.5,3.5)    node (v1)  {$E_{0,N-1}$};
	\path (1.5,3.5)    node (v1)  {$E_{1,N-1}$};
	\path (3.5,3.5)    node (v1)  {$E_{M-1,N-1}$};
	\draw[line width = 1] (-0.1,1) -- (0.1,1);
	\path (-0.3,1) node[rotate=90] (v1) {$h_y$};
	\draw[line width = 1] (-0.1,2) -- (0.1,2);
	\path (-0.3,2) node[rotate=90] (v1) {$2h_y$};
	\draw[line width = 1] (-0.1,3) -- (0.1,3);
	\path (-0.3,3) node[rotate=90] (v1) {$B-h_y$};
	\draw[line width = 2] (-0.1,4) -- (0.1,4);
	\path (-0.3,4) node[rotate=90] (v1) {$B$};
	\path (-0.3,4.8) node[rotate=90] (v1) {$y$};
	\path (4.8,-0.3) node (v1) {$x$};
	\draw[line width = 1] (1,-0.1) -- (1, 0.1);
	\path (1,-0.3) node (v1) {$h_x$};
	\draw[line width = 1] (2,-0.1) -- (2, 0.1);
	\path (2,-0.3) node (v1) {$2h_x$};
	\draw[line width = 1] (3,-0.1) -- (3, 0.1);
	\path (3,-0.3) node (v1) {$A-h_x$};
	\draw[line width = 1] (4,-0.1) -- (4, 0.1);
	\path (4,-0.3) node (v1) {$A$};
\end{tikzpicture}
	\caption{Illustration of the division of $[0,A]\times[0,B]$ into a 
		 regular $M\times N$ grid of rectangular cells.}
	\label{fig:discGrid}
\end{figure}
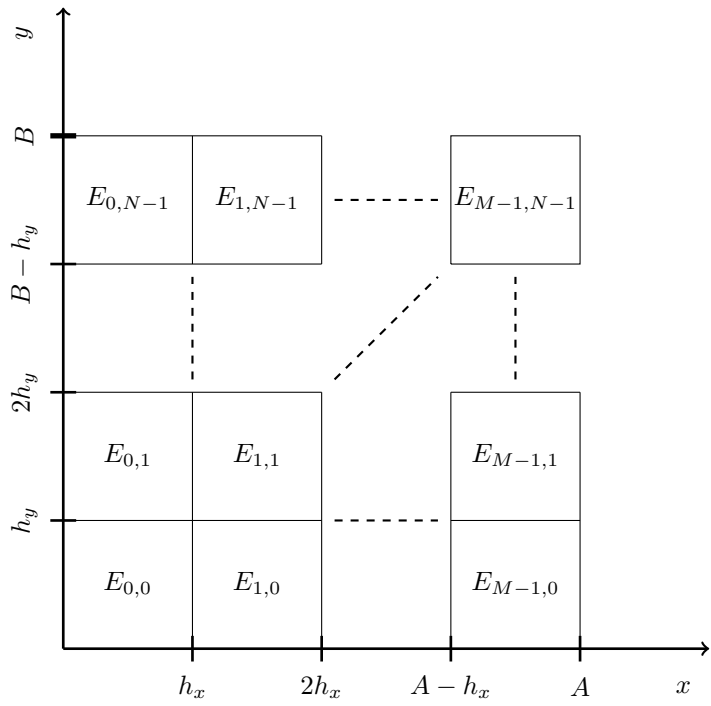

To keep the calculations simple the domain is divided into a regular grid of 
rectangular cells. Use $M$ cells in the $x$-direction and $N$ cells in the 
$y$-direction. Then for each cell the sides parallel to the $x$-axis have 
length $h_x=A/M$ and the sides parallel to the $y$-axis have length 
$h_y = B/N$. Number the cells by $(i,j)$, where $i$ is the column of the cell
(along the $x$-axis) and $j$ is the row of the cell (along the $y$-axis). Call 
the lowest row $0$ and the leftmost column $0$, then cell $(i,j)$ is 
\[
	E_{i,j} = [i h_x,(i+1) h_x] \times [j h_y, (j+1) h_y].
\]
Using this notation the set of cells, $\mathcal{I}$, is given by 
\[
	\mathcal{I} = \{E_{i,j} : i = 0,1,\ldots,M-1, j = 0,1,\ldots, N-1 \}.
\]
Figure~\ref{fig:discGrid} shows an illustration of the discretization of 
$[0,A]\times[0,B]$ into the cells $\mathcal{I}$.

Each cell has four faces, two parallel to the $x$-axis (top and bottom) and 
two parallel to the $y$-axis (left and right). Let the right face, top face, 
left face and bottom face of cell $E_{i,j}$ be denoted $\sigma_{i,j}^{\mathrm{R}}$, 
$\sigma_{i,j}^{\mathrm{T}}$, $\sigma_{i,j}^{\mathrm{L}}$ and $\sigma_{i,j}^{\mathrm{B}}$,
respectively. Additionally, denote by $\sigma(E_{i,j})$ the set of faces
of cell $E_{i,j}$. 

For each cell $E_{i,j}$, $\boldsymbol{s}_{i,j}$ 
gives the centroid of the cell, and  $\boldsymbol{s}_{i+1/2,j}$, 
$\boldsymbol{s}_{i,j+1/2}$, $\boldsymbol{s}_{i-1/2,j}$ 
and $\boldsymbol{s}_{i,j-1/2}$ give the centres of 
the faces of the cell. Due to the periodic boundary conditions, the $i$-index 
and $j$-index in $\boldsymbol{s}_{i,j}$ are modulo $M$ and modulo $N$,
respectively. Figure~\ref{fig:oneCell} shows one cell $E_{i,j}$ with the 
centroid and the faces marked on the figure. 
Further, let $u_{i,j} = u(\boldsymbol{s}_{i,j})$ for each cell and denote the 
area of $E_{i,j}$ by $V_{i,j}$. Since the grid is regular, all $V_{i,j}$ are 
equal to $V = h_x h_y$.

\begin{figure}
	\centering
	\begin{tikzpicture}[]
	\draw[line width=0.8] (0,0) -- (5,0) -- (5,5) -- (0,5) -- (0,0);
	\path (5.8,2.5) node (v1) {$\sigma_{i,j}^{\mathrm{R}}$};
	\path (2.5,5.5) node (v1) {$\sigma_{i,j}^{\mathrm{T}}$};
	\path (-0.8,2.5) node (v1) {$\sigma_{i,j}^{\mathrm{L}}$};
	\path (2.5,-0.5) node (v1) {$\sigma_{i,j}^{\mathrm{B}}$};

	\draw[line width=0.8] (2.4,2.4) -- (2.6,2.6);
	\draw[line width=0.8] (2.4,2.6) -- (2.6,2.4);
	\path (2.5, 2.9) node (v1) {$\boldsymbol{s}_{i,j}$};

	\draw[line width=0.8] (-0.1,2.4) -- (0.1,2.6);
	\draw[line width=0.8] (-0.1,2.6) -- (0.1,2.4);
	\path (0.9, 2.5) node (v1) {$\boldsymbol{s}_{i-1/2,j}$};

	\draw[line width=0.8] (2.4,4.9) -- (2.6,5.1);
	\draw[line width=0.8] (2.4,5.1) -- (2.6,4.9);
	\path (2.5, 4.6) node (v1) {$\boldsymbol{s}_{i,j+1/2}$};
	
	\draw[line width=0.8] (4.9,2.4) -- (5.1,2.6);
	\draw[line width=0.8] (4.9,2.6) -- (5.1,2.4);
	\path (4.1, 2.5) node (v1) {$\boldsymbol{s}_{i+1/2,j}$};

	\draw[line width=0.8] (2.4,-0.1) -- (2.6,0.1);
	\draw[line width=0.8] (2.4,0.1)  -- (2.6,-0.1);
	\path (2.5,0.4) node (v1) {$\boldsymbol{s}_{i,j-1/2}$};

\end{tikzpicture}
	\caption{One cell, $E_{i,j}$, of the discretization with faces 
		 $\sigma_{i,j}^{\mathrm{R}}$, $\sigma_{i,j}^{\mathrm{T}}$, 
		 $\sigma_{i,j}^{\mathrm{L}}$ and $\sigma_{i,j}^{\mathrm{B}}$, centroid 
		 $\boldsymbol{s}_{i,j}$ and centres of the faces 
		 $\boldsymbol{s}_{i-1/2,j}$, $\boldsymbol{s}_{i,j-1/2}$, 
		 $\boldsymbol{s}_{i+1/2,j}$ and $\boldsymbol{s}_{i,j+1/2}$.}
	\label{fig:oneCell}
\end{figure}

To derive the finite volume scheme, begin by integrating 
Equation~\eqref{eq:mainSPDE} over a cell, $E_{i,j}$. This gives
\begin{equation}
	\int_{E_{i,j}} \! \kappa^2(\boldsymbol{s})u(\boldsymbol{s}) \, \mathrm{d}\boldsymbol{s} - \int_{E_{i,j}} \! \nabla\cdot \mathbf{H}(\boldsymbol{s})\nabla u(\boldsymbol{s}) \, \mathrm{d}\boldsymbol{s} = \int_{E_{i,j}} \! \mathcal{W}(\boldsymbol{s}) \, \mathrm{d}\boldsymbol{s},
	\label{eq:fInt}
\end{equation}
where $\mathrm{d}\boldsymbol{s}$ is an area element. 
The integral on the right hand side is 
distributed as a Gaussian variable with mean $0$ and variance $V$ for each 
$(i,j)$~\cite{Adler2007}*{pp.~24--25}. 
Further, the integral on the right hand side is independent for 
different cells, because two different cells can at most share a common
face. Thus Equation~\eqref{eq:fInt} can be written as
\[
	\int_{E_{i,j}} \! \kappa^2(\boldsymbol{s})u(\boldsymbol{s}) \, \mathrm{d}\boldsymbol{s} - \int_{E_{i,j}} \! \nabla\cdot \mathbf{H}(\boldsymbol{s})\nabla u(\boldsymbol{s}) \, \mathrm{d}\boldsymbol{s} = \sqrt{V} z_{i,j},
\]
where $z_{i,j}$ is a standard Gaussian variable for each $(i,j)$ and the 
Gaussian variables are independent.

By the divergence theorem in Equation~\eqref{eq:divThm}, the second integral 
on the left hand side can be written as an integral over the boundary of the 
cell. This results in
\begin{equation}
	\int_{E_{i,j}} \! \kappa^2(\boldsymbol{s})u(\boldsymbol{s}) \, \mathrm{d}\boldsymbol{s} - \oint_{\partial E_{i,j}} \! (\mathbf{H}(\boldsymbol{s}) \nabla u(\boldsymbol{s}))^\mathrm{T} \boldsymbol{n}(\boldsymbol{s}) \, \mathrm{d}\sigma = \sqrt{V}z_{i,j},
	\label{eq:divApplied}
\end{equation}
where $\boldsymbol{n}$ is the exterior normal vector of $\partial E_{i,j}$ 
with respect to $E_{i,j}$ and $\mathrm{d}\sigma$ is a line element. It is 
useful to divide the integral over the boundary in 
Equation~\eqref{eq:divApplied} into integrals over each face, 
\begin{equation}
	\int_{E_{i,j}} \! \kappa^2(\boldsymbol{s})u(\boldsymbol{s}) \, \mathrm{d}\boldsymbol{s} - \left(W_{i,j}^R+W_{i,j}^T+W_{i,j}^L+W_{i,j}^B\right) = \sqrt{V}z_{i,j},
	\label{eq:flux}
\end{equation}
where $W_{i,j}^{\mathrm{dir}} = \int_{\sigma_{i,j}^{\mathrm{dir}}} (\mathbf{H}(\boldsymbol{s})\nabla u(\boldsymbol{s}))^\mathrm{T} \boldsymbol{n}(\boldsymbol{s}) \, \mathrm{d}\sigma$.

The first integral on the left hand side of Equation~\eqref{eq:flux} is 
approximated by 
\begin{equation}
	\int_{E_{i,j}} \! \kappa^2(\boldsymbol{s}) u(\boldsymbol{s}) \, \mathrm{d}\boldsymbol{s} = V \kappa_{i,j}^2 u(\boldsymbol{s}_{i,j}) = V \kappa_{i,j}^2 u_{i,j},
	\label{eq:fApp}
\end{equation}
where $\kappa_{i,j}^2 = \frac{1}{V}\int_{E_{i,j}} \! \kappa^2(\boldsymbol{s}) \, \mathrm{d}\boldsymbol{s}$.
The function $\kappa^2$ is assumed to be continuous and $\kappa_{i,j}^2$ is 
approximated by $\kappa^2(\boldsymbol{s}_{i,j})$.

The second part of Equation~\eqref{eq:flux} requires the approximation of the
surface integral over each face of a given cell. The values of $\mathbf{H}$ are
in general not diagonal, so it is necessary to estimate both components of the 
gradient on each face of the cell. For simplicity, it is assumed that the 
gradient is constant on each face and that it is identically equal to the value
at the centre of the face. On a face parallel to the $y$-axis the estimate of 
the partial derivative with respect to $x$ is simple since the centroid of each
of the cells which share the face have the same $y$-coordinate. The problem is 
the estimate of the partial derivative with respect to $y$. The reverse is true
for the top and bottom face of the cell.

It is important to use a scheme which gives the same estimate of the gradient 
for a given face no matter which of the two neighbouring cells are chosen. For 
the right face of $E_{i,j}$, that is $\sigma_{i,j}^{\mathrm{R}}$, the approximation 
used is
\[
	\frac{\partial}{\partial y} u(\boldsymbol{s}_{i+1/2,j}) \approx \frac{1}{h_y}(u(\boldsymbol{s}_{i+1/2,j+1/2})-u(\boldsymbol{s}_{i+1/2,j-1/2})).
\]
where the values of $u$ at $\boldsymbol{s}_{i+1/2,j+1/2}$ and 
$\boldsymbol{s}_{i+1/2,j-1/2}$ are linearly interpolated from the values at 
the four closest cells. More precisely, because of the regularity of the grid
the mean of the four closest cells are used. This gives
\begin{equation}
	\frac{\partial}{\partial y}u(\boldsymbol{s}_{i+1/2,j}) \approx \frac{1}{4h_y}(u_{i+1,j+1}+u_{i,j+1}-u_{i,j-1}-u_{i+1,j-1}).
	\label{eq:ygrady}
\end{equation}
Note that this formula can be used for the partial derivative with respect to 
$y$ on any face parallel to the $y$-axis by suitably changing the $i$ and $j$ 
indices. The partial derivative with respect to $x$ on a face parallel to the 
$y$-axis can be approximated directly by 
\begin{equation}
	\frac{\partial}{\partial x}u(\boldsymbol{s}_{i+1/2,j}) \approx \frac{1}{h_x} (u_{i+1,j}-u_{i,j}).
	\label{eq:ygradx}
\end{equation}

In more or less exactly the same way the two components of the gradient on the 
top face of cell $E_{i,j}$ can be approximated by
\[
	\frac{\partial}{\partial x}u(\boldsymbol{s}_{i,j+1/2}) \approx \frac{1}{4h_x}(u_{i+1,j+1}+u_{i+1,j}-u_{i-1,j}-u_{i-1,j+1})
\]
and
\[
	\frac{\partial}{\partial y}u(\boldsymbol{s}_{i,j+1/2}) \approx \frac{1}{h_y}(u_{i,j+1}-u_{i,j}).
\]
These approximations can be used on any side parallel to the $x$-axis by 
changing the indices appropriately. 

\begin{table}
	\centering
	\caption{\sf Finite difference schemes for the partial derivative with 
		 respect to $x$ and $y$ at the different faces of cell 
		 $E_{i,j}$.}
	\begin{tabular}{lcc}
		{\bf Face}	    &	$\frac{\partial}{\partial x}u(s)$  &  $\frac{\partial}{\partial y}u(s)$  \\[0.1cm]
		\hline \\[-0.25cm]
		$\sigma_{i,j}^{\mathrm{R}}$  &  $\frac{u_{i+1,j}-u_{i,j}}{h_x}$        & $\frac{u_{i,j+1}+u_{i+1,j+1}-u_{i,j-1}-u_{i+1,j-1}}{4h_y}$ \\[0.2cm]
		$\sigma_{i,j}^{\mathrm{T}}$  & $\frac{u_{i+1,j}+u_{i+1,j+1}-u_{i-1,j}-u_{i-1,j+1}}{4h_x}$  & $\frac{u_{i,j+1}-u_{i,j}}{h_y}$ \\[0.2cm]
		$\sigma_{i,j}^{\mathrm{L}}$  &  $\frac{u_{i,j}-u_{i-1,j}}{h_x}$        & $\frac{u_{i-1,j+1}+u_{i,j+1}-u_{i-1,j-1}-u_{i,j-1}}{4h_y}$ \\[0.2cm]
		$\sigma_{i,j}^{\mathrm{B}}$  & $\frac{u_{i+1,j}+u_{i+1,j-1}-u_{i-1,j-1}-u_{i-1,j}}{4h_x}$  & $\frac{u_{i,j}-u_{i,j-1}}{h_y}$ \\

	\end{tabular}
	\label{table:findiff}
\end{table}

The approximations for the partial derivatives on each face are collected in
Table~\ref{table:findiff}. Using this table one can find the approximations
needed for the second part of Equation~\eqref{eq:flux}.
It is helpful to write 
\[
	W_{i,j}^{\mathrm{dir}} = \int_{\sigma_{i,j}^{\mathrm{dir}}} \! (\mathbf{H}(\boldsymbol{s})\nabla u(\boldsymbol{s}))^\text{T} \boldsymbol{n}(\boldsymbol{s}) \, \mathrm{d}\sigma = \int_{\sigma_{i,j}^{\mathrm{dir}}} \! (\nabla u(\boldsymbol{s}))^\text{T} (\mathbf{H}(\boldsymbol{s}) \boldsymbol{n}(\boldsymbol{s})) \, \mathrm{d}\sigma,
\]
where the symmetry of $\mathbf{H}$ is used to avoid transposing the matrix.
Assuming that the gradient is identically equal to the value at the centre of 
the face, one finds
\[
	W_{i,j}^{\mathrm{dir}} \approx \left(\nabla u(\boldsymbol{c}_{i,j}^{\mathrm{dir}})\right)^\text{T} \int_{\sigma_{i,j}^{\mathrm{dir}}} \! \mathbf{H}(\boldsymbol{s})\boldsymbol{n}(\boldsymbol{s}) \, \mathrm{d}\sigma,
\]
where $\boldsymbol{c}_{i,j}^{\mathrm{dir}}$ is the centre of face $\sigma_{i,j}^{\mathrm{dir}}$.

Since the cells form a regular grid, $\boldsymbol{n}$ is constant on each face.
Let $\mathbf{H}$ be approximated by its value at the centre of the face, then
\begin{equation}
	W_{i,j}^{\mathrm{dir}} \approx m(\sigma_{i,j}^{\mathrm{dir}})\left(\nabla u(\boldsymbol{c}_{i,j}^{\mathrm{dir}})\right)^{\mathrm{T}}\left(\mathbf{H}(\boldsymbol{c}_{i,j}^{\mathrm{dir}})\boldsymbol{n}(\boldsymbol{c}_{i,j}^{\mathrm{dir}})\right),
	\label{eq:Wapp}
\end{equation}
where $m(\sigma_{i,j}^{\mathrm{dir}})$ is the length of the face. Note that 
the length of the face is either $h_x$ or $h_y$ and that the normal vector is
parallel to the $x$-axis or the $y$-axis.

Let 
\[
\mathbf{H}(\boldsymbol{s}) = \begin{bmatrix} 
	H^{11}(\boldsymbol{s}) & H^{12}(\boldsymbol{s}) \\ 
	H^{21}(\boldsymbol{s}) & H^{22}(\boldsymbol{s})
			\end{bmatrix},
\]
then using Table~\ref{table:findiff} one finds the approximations
\begin{eqnarray*}
	\lefteqn{\hat{W}_{i,j}^{\mathrm{R}} =} \\
	& & h_y \left[H^{11}(\boldsymbol{s}_{i+1/2,j})\frac{u_{i+1,j}-u_{i,j}}{h_x}\right]+ \\
	& & h_y\left[H^{21}(\boldsymbol{s}_{i+1/2,j})\frac{u_{i,j+1}+u_{i+1,j+1}-u_{i,j-1}-u_{i+1,j-1}}{4 h_y}\right],
\end{eqnarray*}
\begin{eqnarray*}
	\lefteqn{\hat{W}_{i,j}^{\mathrm{T}} =} \\
	& & h_x \left[H^{12}(\boldsymbol{s}_{i,j+1/2})\frac{u_{i+1,j+1}+u_{i+1,j}-u_{i-1,j+1}-u_{i-1,j}}{4 h_x}\right]+ \\ 
	& & h_x\left[H^{22}(\boldsymbol{s}_{i,j+1/2})\frac{u_{i,j+1}-u_{i,j}}{h_y}\right],
\end{eqnarray*}
\begin{eqnarray*}
	\lefteqn{\hat{W}_{i,j}^{\mathrm{L}} =} \\
	& & h_y \left[H^{11}(\boldsymbol{s}_{i-1/2,j})\frac{u_{i-1,j}-u_{i,j}}{h_x}\right]+ \\
	& & h_y\left[H^{21}(\boldsymbol{s}_{i-1/2,j})\frac{u_{i,j-1}+u_{i-1,j-1}-u_{i-1,j+1}-u_{i,j+1}}{4 h_y}\right]
\end{eqnarray*}
and
\begin{eqnarray*}
	\lefteqn{\hat{W}_{i,j}^{\mathrm{B}} =} \\
	& & h_x \left[H^{12}(\boldsymbol{s}_{i,j-1/2})\frac{u_{i-1,j}+u_{i-1,j-1}-u_{i+1,j}-u_{i+1,j-1}}{4 h_x}\right]+ \\
	& & h_x\left[H^{22}(\boldsymbol{s}_{i,j-1/2})\frac{u_{i,j-1}-u_{i,j}}{h_y}\right].
\end{eqnarray*}
These approximations can be combined with the approximations in 
Equation~\eqref{eq:fApp} and inserted into Equation~\eqref{eq:flux} to give
\[
	V \kappa_{i,j}^2 u_{i,j} - \left(\hat{W}_{i,j}^\mathrm{R}+\hat{W}_{i,j}^\mathrm{T}+\hat{W}_{i,j}^\mathrm{L}+\hat{W}_{i,j}^\mathrm{B}\right) = \sqrt{V} z_{i,j}.
\]
Stacking the variables $u_{i,j}$ row-wise in a vector $\boldsymbol{u}$, that 
is first row $0$, then row $1$ and so on, gives the linear system of equations,
\begin{equation}
	\mathbf{D}_V \mathbf{D}_{\kappa^2}\boldsymbol{u}-\mathbf{A}_{\mathbf{H}} \boldsymbol{u} = \mathbf{D}_V^{1/2}\boldsymbol{z},
	\label{eq:mateq}
\end{equation}
where $\mathbf{D}_V = V \mathbf{I}_{MN}$, 
$\mathbf{D}_{\kappa^2} = \mathrm{diag}(\kappa_{0,0}^2,\ldots,\kappa_{M-1,0}^2,\kappa_{0,1}^2,\ldots,\kappa_{M-1,N-1}^2)$, $\boldsymbol{z} \sim \mathcal{N}_{MN}(\boldsymbol{0}, \mathbf{I}_{MN})$  and $\mathbf{A}_\mathbf{H}$ 
is considered more closely in what follows.

The construction of the matrix $\mathbf{A}_\mathbf{H}$, which depends on the function $\mathbf{H}$, requires 
only that one writes out the sum
\[
	\hat{W}_{i,j}^{\mathrm{R}}+\hat{W}_{i,j}^\mathrm{T}+\hat{W}_{i,j}^\mathrm{L}+\hat{W}_{i,j}^\mathrm{B}
\]
and collects the coefficients of the  different $u_{a,b}$ terms.
This is not difficult, but requires many lines of equations.
Therefore, only the resulting coefficients are
given. Fix $(i,j)$ and consider the equation for cell $E_{i,j}$. For convenience,
let $i_p$ and $i_n$ be the column left and right  of the current column respectively and 
let $j_n$ and $j_p$ be the row above and below the current row respectively. These
rows and columns are $0$-indexed and due to the periodic boundary conditions
one has, for example, that column $0$ is to the right of column $M-1$.
Further, number the rows and columns of the matrix $\mathbf{A}_\mathbf{H}$ from $0$ to $MN-1$.

For row $jM+i$ the coefficient of $u_{i,j}$ itself is given by
\begin{eqnarray*}
	\lefteqn{(\mathbf{A}_\mathbf{H})_{jM+i,jM+i} =} \\
	& & { } -\frac{h_y}{h_x}\left[H^{11}(\boldsymbol{s}_{i+1/2,j})+H^{11}(\boldsymbol{s}_{i-1/2,j})\right] \\
	& & { } -\frac{h_x}{h_y}\left[H^{22}(\boldsymbol{s}_{i,j+1/2})+H^{22}(\boldsymbol{s}_{i,j-1/2})\right].
\end{eqnarray*}
The four closest neighbours have coefficients
\[
	\begin{aligned}
		(\mathbf{A}_\mathbf{H})_{jM+i,jM+i_p} &= \frac{h_y}{h_x}H^{11}(s_{i-1/2,j})-\frac{1}{4}\left[H^{12}(\boldsymbol{s}_{i,j+1/2})-H^{12}(\boldsymbol{s}_{i,j-1/2})\right], \\
		(\mathbf{A}_\mathbf{H})_{jM+i,jM+i_n} &= \frac{h_y}{h_x}H^{11}(\boldsymbol{s}_{i+1/2,j})+\frac{1}{4}\left[H^{12}(\boldsymbol{s}_{i,j+1/2})-H^{12}(\boldsymbol{s}_{i,j-1/2})\right], \\
		(\mathbf{A}_\mathbf{H})_{jM+i,j_nM+i} &= \frac{h_x}{h_y}H^{22}(\boldsymbol{s}_{i,j+1/2})+\frac{1}{4}\left[H^{21}(\boldsymbol{s}_{i+1/2,j})-H^{21}(\boldsymbol{s}_{i-1/2,j})\right], \\
		(\mathbf{A}_\mathbf{H})_{jM+i,j_pM+i} &= \frac{h_x}{h_y}H^{22}(\boldsymbol{s}_{i,j-1/2})-\frac{1}{4}\left[H^{21}(\boldsymbol{s}_{i+1/2,j})-H^{21}(\boldsymbol{s}_{i-1/2,j})\right].
	\end{aligned}
\]
Lastly, the four diagonally closest neighbours have coefficients
\[
	\begin{aligned}
		(\mathbf{A}_\mathbf{H})_{jM+i,j_pM+i_p}  &= +\frac{1}{4}\left[H^{12}(\boldsymbol{s}_{i,j-1/2})+H^{21}(\boldsymbol{s}_{i-1/2,j})\right], \\
		(\mathbf{A}_\mathbf{H})_{jM+i,j_pM+i_n}  &= -\frac{1}{4}\left[H^{12}(\boldsymbol{s}_{i,j-1/2})+H^{21}(\boldsymbol{s}_{i+1/2,j})\right], \\
		(\mathbf{A}_\mathbf{H})_{jM+i,j_nM+i_p}  &= -\frac{1}{4}\left[H^{12}(\boldsymbol{s}_{i,j+1/2})+H^{21}(\boldsymbol{s}_{i-1/2,j})\right], \\
		(\mathbf{A}_\mathbf{H})_{jM+i,j_nM+i_n}  &= +\frac{1}{4}\left[H^{12}(\boldsymbol{s}_{i,j+1/2})+H^{21}(\boldsymbol{s}_{i+1/2,j})\right].
	\end{aligned}
\]
The rest of the elements of row $jM+i$ are $0$.

Based on Equation~\eqref{eq:mateq} one can write 
\[
	\boldsymbol{z} = \mathbf{D}_V^{-1/2}\mathbf{A}\boldsymbol{u},
\]
where $\mathbf{A} = \mathbf{D}_V\mathbf{D}_{\kappa^2}-\mathbf{A}_\mathbf{H}$. 
This gives the joint distribution of $\boldsymbol{u}$,
\begin{align}
	\nonumber	\pi(\boldsymbol{u}) &\propto \pi(\boldsymbol{z}) \propto \exp\left(-\frac{1}{2}\boldsymbol{z}^{\mathrm{T}}\boldsymbol{z}\right) \\
	\nonumber	\pi(\boldsymbol{u}) &\propto \exp\left(-\frac{1}{2}\boldsymbol{u}^{\text{T}}\mathbf{A}^{\mathrm{T}}\mathbf{D}_V^{-1}\mathbf{A}\boldsymbol{u}\right) \\
	\nonumber	\pi(\boldsymbol{u}) &\propto \exp\left(-\frac{1}{2}\boldsymbol{u}^{\text{T}}\mathbf{Q}\boldsymbol{u}\right),
\end{align}
where $\mathbf{Q} = \mathbf{A}^{\mathrm{T}}\mathbf{D}_V^{-1}\mathbf{A}$. This 
is a sparse matrix with a maximum of $25$ non-zero elements on each row, 
corresponding to the point itself, its 8 closest neighbours and the 8 closest 
neighbours of each of the 8 closest neighbours.

\section{Marginal variances with constant coefficients}
\begin{prop}
\label{prop:margVar}
	Let $u$ be a stationary solution of the SPDE 
	\begin{equation}
		\kappa^2 u(x,y) - \nabla\cdot \mathbf{H} \nabla u(x,y) = \mathcal{W}(x,y), \qquad (x,y)\in\mathbb{R}^2,
		\label{chap2:eq:propR2}
	\end{equation}
	where $\mathcal{W}$ is a standard Gaussian white noise process, 
	$\kappa^2 > 0$ is a constant, $\mathbf{H}$ is a positive definite 
	$2\times 2$ matrix and 
	$\nabla = \left(\frac{\partial}{\partial x},\frac{\partial}{\partial y}\right)$.

	Then $u$ has marginal variance
	\[
		\sigma_m^2 = \frac{1}{4\pi \kappa^2 \sqrt{\mathrm{det}(\mathrm{H})}}.
	\]

	\begin{proof}
		Since the solution is stationary, Gaussian white noise is 
		stationary and the SPDE has constant coefficients, the SPDE 
		is acting as a linear filter. Thus one can use spectral theory 
		to find the marginal variance. The transfer function of the 
		SPDE is
		\[
			g(\boldsymbol{w}) = \frac{1}{\kappa^2+\boldsymbol{w}^\mathrm{T}\mathbf{H}\boldsymbol{w}}.
		\]
		Further, the spectral density of a standard Gaussian white 
		noise process on $\mathbb{R}^2$ is identically equal to 
		$1/(2\pi)^2$. It follows that the spectral density of the 
		solution is
		\[
			f_S(\boldsymbol{w}) = \left(\frac{1}{2\pi}\right)^2 \frac{1}{(\kappa^2+\boldsymbol{w}^\mathrm{T}\mathbf{H}\boldsymbol{w})^2}.
		\]

		From the spectral density it is only a matter of integrating 
		the density over $\mathbb{R}^2$,
		\[
			\sigma_m^2 = \int_{\mathbb{R}^2} \! f_S(\boldsymbol{w}) \, \mathrm{d}\boldsymbol{w}. 
		\]
		The matrix $\mathbf{H}$ is (symmetric) positive definite
		and, therefore, has a (symmetric) positive definite square
		root, say $\mathbf{H}^{1/2}$. Use the change of variables
		$\boldsymbol{w} = \kappa\mathbf{H}^{-1/2}\boldsymbol{z}$
		to find
		\begin{align*}
			\sigma_m^2 &= \frac{1}{4\pi^2} \int_{\mathbb{R}^2} \! \frac{1}{(\kappa^2+\kappa^2\boldsymbol{z}^\mathrm{T}\boldsymbol{z})^2} \mathrm{det}(\kappa\mathbf{H}^{-1/2}) \, \mathrm{d}\boldsymbol{z} \\
				   &= \frac{1}{4\pi^2 \kappa^2 \sqrt{\mathrm{det}(\mathbf{H})}}\int_{\mathbb{R}^2} \! \frac{1}{(1+\boldsymbol{z}^\mathrm{T}\boldsymbol{z})^2} \, \mathrm{d}\boldsymbol{z} \\
				   &= \frac{1}{4\pi \kappa^2\sqrt{\mathrm{det}(\mathbf{H})}}.
		\end{align*}
	\end{proof}

\end{prop}


\end{document}